\documentclass[showpacs,aps,pra,floatfix,preprint,superscriptaddress]{revtex4-1}
\usepackage[english]{babel} 
\usepackage[T1]{fontenc} 
\usepackage[utf8]{inputenc} 
\usepackage{lmodern}
\usepackage{microtype}
\usepackage{hyperref} 
\usepackage{amsfonts, vmargin}
\usepackage{amsmath, amssymb,mathrsfs}
\numberwithin{equation}{section} 
\renewcommand{\theequation}{\arabic{section}.\arabic{equation}}
\usepackage{graphicx}
\usepackage{tikz}
\usepackage{epic, eepic}
\usepackage[us]{datetime}
\usepackage{color}
\newcommand{\llangle}{\langle \! \langle} 
\newcommand{\rrangle}{\rangle \! \rangle} 

\newcommand{\G}{{\bf \Gamma}}
\newcommand{\Pb}{{\langle \Phi(t) \vert}}
\newcommand{\Gk}{{ \vert \Gamma(t) \rangle}}
\newcommand{\bx}{{\bf x}}
\newcommand{\bX}{{\bf X}}
\newcommand{\bP}{{\bf P}}
\newcommand{\bp}{{\bf p}}
\newcommand{\by}{{\bf y}}
\newcommand{\balpha}{{\bf \alpha}}
\newcommand{\smeq}{\!=\!}
\newcommand{\terg}{\tau_{\rm erg}}

\definecolor{BLACK}{named}{black}
\definecolor{GREEN}{named}{green}
\newcommand{\sout}[1]{\unskip}

\newcommand{\chkout}[1]{\unskip}

\begin{document}
\title{Quadratic  mean field games}
\author{Denis Ullmo}
\affiliation{LPTMS, CNRS, Univ. Paris-Sud, Universit\'e Paris-Saclay,
  91405 Orsay, France}  
\author{Igor Swiecicki}
\affiliation{LPTMS, CNRS, Univ. Paris-Sud, Universit\'e Paris-Saclay,
  91405 Orsay, France}  
\affiliation{LPTM, CNRS, Univ. Cergy-Pontoise, 95302 Cergy-Pontoise,
  France}
\author{Thierry Gobron} 
\affiliation{LPTM, CNRS, Univ. Cergy-Pontoise, 95302 Cergy-Pontoise,
  France}
\begin{abstract}
{
Mean field games were introduced independently by J-M. Lasry and
P-L. Lions, and by M. Huang, R.P. Malhamé and P. E. Caines, in
order to bring a new approach to optimization problems with a
    large number of interacting agents. The description of such models
    split in two parts, one describing the evolution of the density of
    players in some parameter space, the other the value of a cost
    functional each player tries to minimize for himself, anticipating
    on the rational behavior of the others.

Quadratic Mean Field Games form a particular class among these
systems, in which the dynamics of each player is governed by a
controlled Langevin equation with an associated cost functional
quadratic in the control parameter. In such cases, there exists a deep
relationship with the non-linear Schrödinger equation in imaginary
time, connexion which lead to effective approximation schemes as well
as a better understanding of the behavior of Mean Field Games.

The aim of this paper is to serve as an introduction to Quadratic Mean
Field Games and their connexion with the non-linear Schrödinger
equation, providing to physicists a good entry point into this new and
exciting field.  }
\end{abstract}
\date{\today}
\maketitle
\ifdef{\sout}{\renewcommand{\sout}[1]{\unskip}}{\newcommand{\sout}[1]{\unskip}}
 

\section{Introduction}

Differential Games represent a field of mathematics at the frontier
between optimization problems and Game Theory. It relates to
optimization problems in the sense that they model socio-economic
phenomena in which agents have control of some parameters (e.g.\ their
velocity, or the amount of resource they dedicate to some goal),
which can be modified in the course of time in order to minimize some
given cost functional.  It relates also to Game Theory because the
cost function of a given agent is assumed to depend not only on his
(or her) own state but also on the one of the other agents involved,
implying a strategic approach to the parameter choices.

When the number of agents becomes large, differential games may become
technically very complex. Practical applications for a large number of
players are therefore extremely difficult to implement
\cite{KhanSun}. In some aspects, this situation is reminiscent of the
problems encountered by physicists when considering systems with a
large number of interacting components, for which an exact treatment
is in most of the cases intractable, but where a ``mean field''
approximation provides in many circumstances a very decent
description.  Owing explicitely to this source of inspiration,
P-L.~Lions and J-M.~Lasry
\cite{LasryLions2006-1,LasryLions2006-2,LasryLions2007}, and
independently M.Huang, R.P. Malham\'e and P. E. Caines
\cite{Huang2006}, introduced a ``mean field'' approximation scheme
adapted to these differential games.  The key idea at the root of Mean
Field Games (MFG in the following), is that the very complexity
associated with a large number of players allows  for a drastic
simplification when considering
that a given agent is not really sensitive to the individual
choices of every agent, but only to an averaged quantity (the
mean field) which aggregates the decisions made by the other
participants to the game.

Since the first articles in 2005-2006, Mean Field Games has developped
into a very active field of research and undergone a rapid growth in
several directions~: On the more formal side, important results have
been obtained on the existence and uniqueness of the associated PDEs
(\cite{GomesSaude2014,Cardaliaguet-course}) or on the convergence of a
many player game to its mean field version
(\cite{CarmonaDelarue2013,BensoussanBook,Cardialaguet2015}). Significant
work has been also done in order to find numerical schemes able to
handle this kind of problems
(\cite{Achdou2012,Gueant2011,LachapelleWolfram2011}). More recently,
application oriented studies have been performed in the context of
finance (\cite{Lachapelle2016,LasryLions2007,Cardialaguet2017}),
economic problems (\cite{GueantLasryLions2010,Achdou2014,Achdou2016}),
or engineering \cite{WirelessNetwork,KizilkaleMalhame2015}.

This recent period is thus marked by important results coming from the
mathematical and engineering science communities, with more recent
contributions from economists, but very little involvement from
physicists.  

One important goal of this review is to demonstrate that this does
  not have to be the case, and that physicists can bring an
  interesting and important point of view to the study of Mean Field
  Games.  Indeed, although the MFG equations allow for a drastic
  simplification of the original problem, they are still very
  difficult to analyze. Very few exact solutions exist, mainly in the
  static case or in very simplified settings
  \cite{Gueant2009,Bardi2012,LaguzetTurinici2015,Gomes2015,Swiecicki-PhysicaA-2016}.
  Furthermore the numerical schemes that have been developed, in spite
  of their quantitative accuracy, do not necessarily provide a
  complete understanding of the mechanism at work in Mean Field
  Games. A physicist's approach to MFG, in which ones develops
  concurrently both an intuition of the qualitative behaviors and
  quantitatively accurate approximation schemes, is therefore very
  much needed.
 
  Thus, rather than an exhausting review of existing mathematical
  results, the aim of this paper is to provide a good {\em entry
    point} for physicists to this fascinating new and rapidly
  developing field\ of Mean Field Games.  For a recent survey of
  mathematical results, the reader is referred to the existing reviews
  \cite{GueantLasryLions2010,GomesSaude2014,Cardaliaguet-course} and
  books \cite{BensoussanBook,CarmonaDelarueBookI,CarmonaDelarueBookII}
  covering in very much details the results obtained by the
  mathematical community.  In this paper, we shall on the other hand
  focus on a particular class of Mean Field Games, the so-called
``{\em quadratic} Mean Field Games'' which is still representative of
the subject at large, and has the peculiarity of exhibiting a direct,
formal, and deep relationship with the {\em non-linear Schr\"odinger
  equation}.  This connection brings a wide set of existing tools to
this field, which because of the long history of the non-linear
  Schr\"odinger equation in physics are both extremely familiar to
physicists and lead to very significant progresses in the
comprehension of the solutions of Mean Field Games models.

The body of this article is divided into five parts. In
section~\ref{sec:review}, we introduce in details the generic form of
Mean Field Games, and present some of its extensions and
applications. This section will be the only one that genuinely qualify
as a [bird's-eye] survey of the mathematical
literature.

The rest of the article is
devoted to the study of the behavior of the quadratic mean field
games, stressing in particular how tools which are well known to
  the physics community can be used effectively to understand the
  corresponding models.  In section~\ref{sec:SchrodingerApproach}, we
derive the connection between quadratic MFG systems and the non linear
Schr\"odinger equation. As a first application, we present a few exact
relations satisfied by some of the MFG statistics and revisit the
stationary problem. We also derive exact
solutions of ``soliton''-type for some specific MFG problems, part
  of them being previously known, and some others new. In the two
following sections, we study thoroughly the case of strong
interactions, both in stationary and 
  non-stationary settings.  Section~\ref{sec:strong} presents the
main methods and results;  in
section~\ref{sec:more-complicated} we extend these results in various
ways (higher dimensions, general initial conditions, etc..).
Finally, in section~\ref{sec:perturbative} we consider the case
of weak interactions that can be  cast into the same framework.



\section{Mean Field Games}
\label{sec:review}

In this section, we first describe in some details the construction of
Mean Field Games in the form in which they were originally introduced
(\cite{LasryLions2006-1,  LasryLions2006-2}), and  review
some generalizations, as well as some example of applications.

\subsection{The ``Mean Field Games'' Paradigm}
\label{sec:basic}

In its original form, a MFG problem can be set as follow.  As a
starting point, we consider a differential game with $N$ players (or
agents). Their individual states are described by a continuous
variable ${\bf X}^i \in \mathbb{R}^d$, $i=1 \dots N$, which, depending
on the context, may represent a physical position, the amount of
resources owned by a company, the house temperature in a network of
controlled heaters, etc..  These state variables 
  evolve according to some controlled dynamics, which
here is assumed to  be described a linear Langevin equation,
\begin{equation} \label{eq:Langevin}
d {\bf X}^i_t = {\bf a}^i_t dt +  \sigma d{\bf W}^i_t\\
\end{equation}
with initial conditions ${\bf X}^i_t={\bf x}^i_0 $. For simplicity,
$\sigma$ is chosen here to be a constant, each of the $d$ components
of ${\bf W}^i$ is an independent white noise of variance one, and the
control parameter is the velocity ${\bf a}^i_t$. This control is
adjusted in time  by the agent $i$ in order to minimize a
cost functional over a time interval $[0,T]$
\begin{equation}\label{eq:cost}
c[{\bf a^i}]({\bf x}^i_t,t) =    
  \llangle \displaystyle \int_{t}^T \left( L({\bf X}^i_\tau,{\bf a}^i_\tau)-
  \tilde V[m_{\tau}]({\bf X}^i_\tau) \right) d\tau \rrangle_{\rm noise}  
  +\llangle c_T({\bf X}^i_T) \rrangle_{\rm noise} \; .
\end{equation}
The cost functional is an average over all realizations of the noise
for an agent starting at ${\bf x}^i_t$ at time $t$ and represents
the rational expectations of each player. In differential games, it is
made of two parts: a time integral over a ``running cost'', here
$L({\bf x},{\bf a})-\tilde V[m]({\bf x})$, and a ``final cost'',
$c_T({\bf x})$, depending on the state of the agent at the end of the
optimization period $T$.  The running cost here splits into two terms,
a ``free Lagrangian part'', $L({\bf x},{\bf a})$, depending only on
the agent's state $\bf x$ and control $\bf a$, and a ``potential''
term, $-V[m_t]({\bf x})$, which takes into account the interaction
between agents and is a functional of the empirical density of agents
$m_t$ in the state space,
\begin{equation}
m_t({\bf x}) = \frac{1}{N}
\sum_j \delta ({\bf x} - {\bf X}^j(t)) \; .
\end{equation}

Implicitly, we have assumed that all agents have an identical
behavior in the sense that they may differ only by their initial
conditions and the subsequent choices of control parameters.
Finally, as usual in control theory \cite{bertsekas2012} one
  introduces a value function which is defined as the minimum, over
  all controls, of the cost function, given the initial condition
  ${\bf x}$ at time $t$. It is independent on the agent label $i$ and
  reads
\begin{equation}\label{eq:value}
u_t({\bf x}) \equiv \min_{{\bf a}}  c[{\bf a}]({\bf x},t) \; .
\end{equation}

Up to now, we have only described a continuous differential game, and
it has to be made clear that even in the above simplified setting,
there is no hope to solve this system beyond a very few number of
players. The aim of Mean Field Games is to provide a frame in which
such a system can be analyzed in the limit of a large number of
players.  One thus renounces
to follow each agent individually, but rather describes the system
at a statistical level through the density of agents
$m_t({\bf x})$.

Now, the essential assumption of Mean Field Games is that the density
of agents becomes deterministic in the large $N$ limit.  As a
consequence, like in standard mean field theories, for a given time
dependent density $m_t$, the trajectories of the different players
decouple, and the problem reduces to an optimization problem for a
single agent.  In particular, using Ito's calculus, the value function
for each player can be shown to be solution of the following
Hamilton-Jacobi-Bellman equation (we give a brief sketch of the
derivation in appendix~\ref{app:HJB})
\begin{equation}\label{eq:hjb}
\left\{
\begin{aligned}
	 &\partial_t u_t(\bx) +H(\bx,\pmb \nabla u_t(\bx))+
         \frac{\sigma^2}{2}  \Delta u_t(\bx) = 
        \tilde V[m_t](\bx)\\
 &u_T(\bx)=c_T(\bx) 
 \end{aligned}
 \right.
 \end{equation}
 where $H(\bx,\bp) \equiv \inf_{\pmb\alpha} \left(
   L(\bx,\pmb\alpha)+\bp .\pmb \alpha) 
\right)$.

The optimal control is then the value of $\bf a$ which realizes
  the above infimum, namely $\bf a= \bf a^*_t(\bx)=
\frac{\partial H}{\partial \bp}(\bx,\nabla 
u_t(\bx))$. Substituting into the Langevin equation
  \eqref{eq:Langevin},  the {\em
  optimized} agent density $m_t$ then evolves from an initial condition
$m^0$ according to the  Kolmogorov equation 
\cite{risken2012}
\begin{equation}
	\left\{
	\begin{aligned} 
	&\partial_t m_t(\bx) + \pmb \nabla . ( m_t(\bx)\,{\bf a}^*_t(\bx) )  -
\frac{\sigma^2}{2} \Delta m_t(\bx) =0 \label{eq:kol}\\
         &m_0(\bx)=m^0(\bx)
	\end{aligned}
	\right. \; .
\end{equation}	

By consistency, the two agent densities (i.e.\ the one used in the
  optimization leading to Eq.~\eqref{eq:hjb} and the one resulting
  from Eq.~\eqref{eq:kol}) need to coincide and we are left with the
  set of partial differential equations \eqref{eq:hjb} and
  \eqref{eq:kol} which are coupled together through the terms
  $\tilde V[m_t]$ and ${\bf a}^*_t$. They are both of diffusion type,
  but respectively backward and forward in time. Together, they form
  the system of equations defining the Mean field
    Game which is expected to describe the large $N$ limit of our
    initial differential game \cite{LasryLions2006-2}.

    In the limit of a very large optimization period $T \to \infty$,
    this system of equation has a remarkable property, proven under
    some specific conditions by P.~Cardialaguet et al.\
    \cite{Cardaliaguet2013}.  In a wide time span, when sufficiently
    far from both limits, $t = 0$ and $t = T$, the system stays in a
    permanent regime where the solution $(m_t(\bx), u_t(\bx))$ of the
    MFG system remains well approximated by the
    $(m^e(\bx), u^e(\bx)e^{\lambda_e t})$, where the couple
    $(m^e(\bx), u^e(\bx))$ is solution of an analog ergodic problem
  (we assume here its existence and unicity)
\begin{equation}\label{eq:MFG-erg}
	\left\{
	\begin{aligned} 
	&- \lambda^e +H(\bx,\pmb \nabla u^e(x))+ \frac{\sigma^2}{2} \Delta u^e(x) =
        \tilde V[m^e](\bx) \\
	&\pmb \nabla(m^e(x) {\bf a}^e(x) )- \frac{\sigma^2}{2} \Delta m^e(x) = 0 \\
	\end{aligned}
	\right. \; ,
\end{equation}			
where
$ H(\bx,\bp)= \inf_{\balpha} \left( L(\bx,\balpha)+\bp \cdot \balpha)
\right)$,
$a^e(x)=\frac{\partial H}{\partial {\bf p}} (\bx,\pmb\nabla u^e(\bx))$
and $\lambda^e$ an appropriately chosen constant.  This system of
equations is referred to as the {\em stationary} Mean Field Games
system of equations.  In the sequels, we will give a transparent and
intuitive interpretation of what is this ergodic solution and how it
is approached in this long time horizon regime.

\subsection{  Overview of recent generalizations. }
\label{sec:extensionI}

The class of Mean Field Games described in the previous subsection
includes already a rich variety of models which behaviors are far from
being fully understood.  In the last few years, very significant
effort have been made to extend the Mean Field Games approach beyond
this original class of problems, or to consider them from a rather
different point of view.  These efforts reflects in part the necessity
to approach a more realistic description of economic or social (or
other) questions, which imposes to relax somehow the restrictive
hypothesis originally assumed in the original formulation described
above.

As we have already stressed, the behavior of even the simplest Mean
Field Games model are very poorly understood (in the sense a
physicist uses for this word), and we shall not address in any details
 these extensions here. Starting from
section \ref{sec:SchrodingerApproach}, we shall actually further limit
our study to a particular subclass of these Mean Field Games (namely
the {\em quadratic} MFG), which, as we shall see, provide a nice entry
point to the field for physicists.  We shall however in this
subsection and the following one provide a brief survey of 
 the main existing lines of research. We refer the interested reader
to recent mathematical review \cite{GomesSaude2014} and monographs \cite{CarmonaDelarueBookI, CarmonaDelarueBookII}.

\subsubsection*{Generalized cost functions}

A very natural way to generalize the basic MFG of
section~\ref{sec:basic} is to enlarge the set of admissible cost
functions.  A first obvious step, actually already taken in the first
papers of Lasry and Lions \cite{LasryLions2006-1,LasryLions2006-2}, is
to assume that the functional $\tilde V$ in Eq.~\eqref{eq:cost} may
also have an explicit dependence in time, reading thus
$\tilde V[m_\tau](\bX^i_\tau,\tau)$.

There are furthermore some situation where the running cost cannot be
written as the sum of a ``free Lagrangian term''
$L({\bf x}\tau,{\bf a}\tau)$ depending only on the state and control
variable of a given agent and a ``potential'' term
$ \tilde V[m_{\tau}]({\bf x})$ which describe its interaction with the
distribution of all agents. This is the case, for instance when
considering pedestrian flow \cite{LachapelleWolfram2011}, where
congestion effects need to be taken into account.  Note that
    congestion here does not means that agents 
avoid crowded places (which in any case would be taken into account by
a proper choice of the functional $\tilde V$), but that
being at a location where the density of agents is high is
    making more costly the use of a large velocity.  In that case, it
is therefore necessary to introduce a term coupling directly the
control parameter with the agent densities.  A particular model for the
congestion phenomena  where the running cost contains  a term proportional to
\[
(m_\tau + \mu)^\gamma |{\bf a}^i_\tau|^\frac{\beta}{\beta-1} 
\]
(with $\mu, \gamma \ge 0$ and $\beta \in (1,2]$)
have been discussed by Lions in \cite{Lions-CollegeClasse} and more
recently by Adchou and Porretta \cite{Achdou2018}.

In another context, Gomes et al.\ \cite{Gomes2014} have also
introduced, and analyzed in the stationary case, a model in which the
cost function of a given player is affected not only by the
distribution of the other players state variable, but also by the
value of their {\em control variables}.  This kind of models arise
naturally in the context of Mean Field Game model of trade crowding,
and have been coined {\em Mean Field Games of Control} by Cardialaguet
and Lehalle \cite{Cardialaguet2017}.

\subsubsection*{Different kinds of players}

One simplifying assumption in the basic MFG of section~\ref{sec:basic}
is that all the player are essentially identical (exchangeable), and
therefore distinguish themselves only through the value of their state
variable.

A natural extension within the Mean Field game hypothesis is to
consider that there exists different groups of agents, $G^{(k)}$,
$k = 1,\cdots,K$, each being characterized by a specific cost function
which may also depend separately on the partial densities
\[
m^{(k)}(\bx) = \frac{1}{N} \sum_{i \in G^{(k)}}
\delta(\bx-\bX^i)
\]
and not only on the total density $m(\bx) = \sum_k m^{(k)}(\bx)$. In
such cases, each agent has to minimize a cost functional similar to
Eq.~\eqref{eq:cost}, but specific to its group, in which the
``potential'' $\tilde V[m]({\bf x})$ is replaced by
$\tilde V^{(k)}[m^{(1)}, \cdots , m^{(K)}]({\bf x})$.

One example of these Mean Field Games involving more than one kind of
population is the model of segregation studied by Adchou et al.\
\cite{Achdou2017}, which is in some sense the analogue of the model
introduce by Schelling in 1971 to study segregation effects in the
United States cities \cite{Schelling1971}.

If  the introduction of different kinds of ``small players'' may lead to rather
new kinds of behavior, it
however does not lead to a very significant conceptual change in the
general  theory of Mean Field Games.  This aspect is however rather
different when the  group(s) of small players interact (strategically) with
one or many ``big players'' \cite{CarmonaZhu2016}, a situation which
is often encountered when modeling  financial markets.  For instance,
in \cite{Lachapelle2016}, Lachapelle et al.\ considered the question
of the price formation in a financial market where a small number of
``institutional investors'' cohabit with a large number of high
frequency traders.  In this context, the high frequency traders sell
or buy only small quantities and only the coherent action of a large
$(O(1))$ fraction of them can affect the outcome.  On the other hand
the institutional investors can buy or sell very large quantities that
can have a significant impact on their own.

In such circumstance, one can still consider that the mean field
hypothesis is still valid for the population of small players.
However, the situation is different for the big players as the
fully stochastic nature of their individual evolution need to be taken
into account.  This eventually leads to a mixed description, with a
mean field game coupled to the stochastic differential game for a
small number of (big) players.

\subsubsection*{Mean Field Games on Graphs}

Another extension of the MFG concept is to apply it to cases where the
state space is discrete rather that continuous.  This changes rather
significantly the structure of the Mean Field Game since the Fokker
Planck equation is replaced by a set of rate equations, and similarly
the Hamilton-Jacobi-Bellman equation by discrete Bellman equations.
The approach to such models is thus technically rather different but
may provide a useful simplified setting when analyzing a particular
qualitative effect, such as for instance congestion \cite{Gueant2015}.
Furthermore, some economic or social problem naturally lead to such
graph-MFG model.  In health science for instance, they provide a
natural setting for a game theoretical version of the SIR
(Susceptible/Infected/Recovered) model for spread of disease, which
have been investigated in details by Laguzet and Turinici
\cite{LaguzetTurinici2015}.

\subsection{Probabilistic approach and the Master Equation}

All  the Mean Field Game extensions  described in the previous
subsection involve non-trivial changes with respect to the original
formulation of section~\ref{sec:basic}. However, {essentially all of them
suppose   that the evolution of the
density of (small) players is described by a {\em deterministic}
equation and that its fluctuations can be neglected.

However, a number of circumstances require to take into account the
existence of stochastic effects, which may survive even in the limit
of infinitely many players.  For instance, this is the case in
presence of a common noise \cite{CarmonaLackerDelarueLacker2016} or
when one has to take into account the stochastic dynamics of big
players \cite{CarmonaZhu2016}.  Moreover, such an extension is also
needed when one wants to relate an $N$-player game with its MFG
counterpart in the $N \to \infty$ limit.

In these cases, it is not possible to assume that the density of
agents can be replaced by its average value, and its stochastic nature
needs to be taken into account. At an heuristic level this can be
understood as implying that the system of coupled forward
Fokker-Planck equation and backward Hamilton-Jacobi-Bellman equation
Eqs.~\eqref{eq:hjb}-\eqref{eq:kol} have to be replaced by a system of
coupled {\em stochastic} Fokker-Planck equation and {\em stochastic}
backward Hamilton-Jacobi-Bellman equation.  In some instances, such as
the case of a large population in a random environment, one can
actually construct, and deal with, the resulting stochastic mean
field game\cite{CarmonaDelarue2014}. However, for most cases, it is
necessary to reformulate the Mean Field Games approach to take into
account the full complexity of these models.  

This means that one does not rely any more on an Hamilton Jacobi
  Bellman equation, which suppose that the time dependent probability
  distribution of player is fixed, but rather consider 
    this distribution as a variable to be fixed in the optimization
  process itself. In \cite{CarmonaDelarue2013}, Carmona and Delarue
  have initiated a purely probabilistic description of Mean Field
    Games, which is an alternative approach to the one based on PDEs
    introduced by Lasry and Lions.  This probabilistic approach
    provides one way to tackle the difficulty implied by the
    stochasticity of the distribution of players
    \cite{CarmonaDelarueBookI,CarmonaLacker2015}. 
  Another framework for this reformulation has been introduced by
P.-L. Lions and called the ``Master Equation'' approach
\cite{Lions-CollegeClasse}. Its basic ingredient consists in writing
the value function $u$ (cf Eq.~\eqref{eq:value}) as a function of
time, state variable and {\em the full} distribution $m$. This lead to
a single differential equation for $u$ which is of second order in $m$
\cite{Cardialaguet2015, Bensoussan2015}.

This Master Equation is a very powerful and sophisticated  tool, that
for obvious reason we shall not describe in any details here, but
it represent presumably the most active direction in the development
of MFG in the mathematical community
\cite{Lions-CollegeClasse,CarmonaDelarue2014,Bensoussan2015,CarmonaLackerDelarueLacker2016,Bensoussan2017,CarmonaDelarueBookII}.

\section{Quadratic Mean Field Games : Schr\"odinger approach}
\label{sec:SchrodingerApproach}

\subsection{ Quadratic Mean Field Games}

In the rest of this paper, we shall restrict our attention to {\sl
  quadratic} Mean Field Games, which are defined by the fact that
the ``free Lagrangian'' part of the running cost has a quadratic
dependence in the control:
\begin{equation}
L({\bf X},{\bf a}) = \frac{1}{2} \mu {\bf   a}^2
  \end{equation} 
  This class of MFG will be shown to admit a mapping to a non linear
  Schr\"odinger Equation, which leads to an almost complete
  description of their behavior.

In addition, we will assume that the potential $\tilde
V[m](\bx)$ can be written as the sum of two terms
\begin{equation}\label{eq:potential}
 \tilde V[m](\bx) = U_0(\bx) + V[m](\bx) 
 \end{equation}
 where $U_0(\bx)$ is an ``external potential'' which depends only on
 the state $\bx$ of the agent while $V[m](\bx)$ describes the interactions between agents and is 
 invariant under  simultaneous  translation of both $\bx$ and $m(\cdot)$.  The simplest form
 that we will consider for this interaction will be linear and local,
\begin{equation}\label{eq:SRLInt}
	V[m](\bx) =  g \,m(\bx)  \; 
\end{equation}
where $g>0$ corresponds to attractive interactions.
We shall  also consider non local interactions
\begin{equation}\label{eq:NLocInt}
	V[m](\bx) = \int d \by \, \kappa (\bx-\by) m(\by)  \; ,
\end{equation}
 and non linear ones
\begin{equation}\label{eq:NLinInt}
	V[m](\bx) = f[m(\bx)]   \; ,
\end{equation}
both admitting the simplest form, Eq.~\eqref{eq:SRLInt} as a
particular case, with, respectively,
$\kappa (\bx-\by) = g \, \delta(\bx-\by)$ and $f(m) = g \, m$.

In the context of crowd dynamics, $U_0(\bx)$ would represent the
preference of an agent for a given position $\bx$, whereas
the term $V[m](\bx)$ takes into account his preference or aversion for crowded places.
In this paper we will limit our study to the attractive case (e.g.\ $g
\geq 0$ in  Equation \eqref{eq:SRLInt}). Two limiting regimes will
be of particular interest: the case of strong interactions dominated
by $V[m](\bx)$ and the case of weak interactions in which $U_0(\bx)$
is the larger term.

To summarise, we consider a set of $N$ agents, whose individual states
at time $t$ are described by continuous variables
${\bf X}^i_t \in \mathbb{R}^d$, which evolves through a controlled
linear Langevin dynamics
\begin{equation}
	d {\bf X}^i_t = {\bf a}^i_t dt + \sigma d {\bf W}^i_t \; ,
\end{equation}
where $\sigma>0$ is a constant, the components of ${\bf W}^i$ are
independent white noises of variance $1$ and ${\bf a}^i$ is the control
chosen by agent $i$ to minimize the cost functional
\begin{equation} \label{eq:cost3}
c[{\bf a}^i]({\bf x}^i_t,t) =   
 \llangle  \int_{t}^T \left( \frac{\mu }{2 }\|{\bf a}^i_\tau\|^2 -
  \tilde V[m_{\tau}]({\bf X}^i_\tau) \right) d\tau  \rrangle_{\rm noise} 
  + \llangle c_T({\bf X}^i_T)  \rrangle_{\rm noise} 
\end{equation}
where $\tilde V[m](\bx)$ is a functional of the density $m$.  In this
setting, the optimal control is ${\bf a}^*_t(\bx)= - \frac{1}{\mu}
  \pmb\nabla u(\bx,t)$ with $u(\bx,t)$ the value function
\eqref{eq:value} and the MFG system \eqref{eq:hjb}--\eqref{eq:kol}
writes here 
\begin{align}
    & \partial_t u_t(\bx)  - \frac{1}{2\mu} \|\pmb \nabla u_t(\bx)\|^2  + 
    \frac{\sigma^2}{2} \Delta \, u_t(\bx) =\tilde V[m_t](\bx)
    \; \label{eq:hjb2}
     \\
    &\partial_t m_t(\bx)  -\frac{1}{\mu} \pmb \nabla . (m_t(\bx) \pmb\nabla u_t(\bx) )  -
    \frac{\sigma^2}{2}\Delta\, m_t(\bx) =0\label{eq:kol2}
\end{align}
 with, respectively, final and initial conditions,
$u_T(\bx)=c_T(\bx)$ and $m_0(\bx)=m^0(\bx)$.

In the following we introduce first a change of variables which shows
that this system of equations is equivalent to a Schr\"odinger
Equation in imaginary time, and the related formalism which we will
use in the rest of this work. We also briefly review three solvable
models: non-interacting agents, interactions in the absence of an
external potential, and quadratic potential.  These models are
interesting in their own rights, but they may also serve as reference
models in perturbative approaches studied in the next sections.

\subsection{Schr\"odinger formalism}

As a first step, we use the well known fact that the  Hamilton-Jacobi-Bellman equation for the value
function $u(\bx,t)$ in \eqref{eq:hjb2} can be
cast into a standard heat equation using a Cole-Hopf transformation
\cite{ColeHopf}  
\begin{equation} \label{eq:Phi-def}
\Phi(\bx,t)=\exp \left( -u_t(\bx)   / \mu\sigma^2 \right) \; .
\end{equation}
The new variable $\Phi(\bx,t)$ obeys a time-backwards diffusion
equation, 
\begin{equation} \label{eq:NLS_Phi}
 - \mu \sigma^2 \, \partial_t \Phi(\bx,t) = \frac{\mu \sigma^4}{2}
      \Delta \Phi(\bx,t) +\tilde V[m_t](\bx)
      \Phi(\bx,t) \; ,
\end{equation}	
with the final condition $\Phi(\bx, T) = \exp(-c_T(\bx)/\mu
\sigma^2)$. Note that it follows from equation \eqref{eq:NLS_Phi}
that 
$\Phi({\bf \cdot, \cdot})>0$  as soon as $\Phi(\bx,T)>0$ everywhere.

The  next step is a change of variables for the density $m_t(\bx)$  \cite{Gueant2011} 
  \begin{equation} \label{eq:Gamma}
 \Gamma(\bx,t) = \frac{m_t(\bx)}{\Phi(\bx,t)} \; .
\end{equation}	
This  second variable now follows a similar heat equation, but forward in time,
\begin{equation} \label{eq:NLS_Gamma}
   \mu \sigma^2 \, \partial_t \Gamma(\bx,t) = \frac{\mu
        \sigma^4}{2} \Delta \Gamma(\bx,t) + \tilde V[m_t](\bx)
      \Gamma (\bx,t) \; ,
  \end{equation}
with the initial condition $\Gamma(\bx,0) ={m^0(\bx)}/{\phi(\bx,0)}$.

Under these transformations, the MFG system has been recast in a pair
of non-linear heat equations, differing only by the sign on the
left hand side and by their asymmetric boundary conditions.

Let us now consider the scalar nonlinear Schr\"odinger equation
describing the quantum evolution of a wave amplitude in a reversed
potential $-\tilde V[\rho]$, 
\begin{equation}\label{eq:NLS}
{\it i} \hbar \; 
\partial_t\psi(\bx,t)
=-\frac{\hbar^2}{2 \mu}
\Delta \psi(\bx,t)  -
\tilde V [\rho](\bx) \psi(\bx,t) \; ,
\end{equation}
with $\rho \equiv \psi^* \psi$ and $\int_{\mathbb R^d} \rho(x)
  =1$.  We note that Eq.~\eqref{eq:NLS} and its complex conjugate are
equivalent to Eqs.~\eqref{eq:NLS_Phi}--\eqref{eq:NLS_Gamma} under the
formal correspondence $\mu \sigma^2 \to \hbar $, $\phi(\bx,t) \to
\psi(\bx, i t)$ and $\Gamma(\bx,t) \to \left[\psi(\bx, i t)\right]$.

 Furthermore the ergodic system
\eqref{eq:MFG-erg},  reads  here
\begin{align} 
	&- \lambda^e  -  \frac{1}{2\mu} {\| \nabla u^e(\bx) \|}^2  + 
	\frac{\sigma^2}{2} \Delta  u^e(\bx)  = \tilde V[m^e](\bx)
	 \; \label{eq:HJBquadra_erg} \\
	& \frac{1}{\mu} \nabla ( m^e(\bx) \nabla u^e(\bx))  +
	\frac{\sigma^2}{2} \Delta  m^e(\bx) = 0 \; . \label{eq:KMGquadra_erg}
\end{align}
When  considering the new variables,  $\Phi^e=\exp \left( - u^e / \mu
  \sigma^2) \right)$ and 
$\G^e= {m^e}/{\Phi^e}$, we see that both  have to follow the same equation 
\begin{equation}   
		\lambda^e \psi^e =  -\frac{ \mu \sigma^4}{2} \Delta  \psi^e 
                     - \tilde V[m^e](\bx)  \psi^e \; ,
              \label{eq:NLSergo}
\end{equation}
with either $\psi^e = \Phi^e(\bx)$ or $\psi^e = \G^e$.  In this 
context, the connection with the nonlinear Schr\"odinger equation
\eqref{eq:NLS} appears more clearly since if $\psi^e(\bx)$  is a solution of 
Eq.~\eqref{eq:NLSergo}, then the two time dependent functions
\begin{align} 
	\phi(\bx,t) &= \exp\bigl\{+\frac{1}{\mu \sigma^2} \lambda^e
        t\bigr\} \psi^e(\bx)\\ 
	\Gamma(\bx,t)& = \exp\bigl\{- \frac{1}{\mu \sigma^2}\lambda^e
        t\bigr\} \psi^e(\bx) 
           \; ,
\end{align}
are solutions of, respectively
Eqs.~\eqref{eq:NLS_Phi}-\eqref{eq:NLS_Gamma}, and simultaneously
  both solutions of  Eq.~\eqref{eq:NLSergo},  in the very
same way as
\[
	\psi(\bx,t)=\exp\bigl\{- \frac{i }{\hbar} \lambda^e t\bigr\}\psi^e(\bx) 
\]
is solution of both Eq.~\eqref{eq:NLS} and Eq.~\eqref{eq:NLSergo}.

The non linear Schr\"odinger equation has been used for decades to
describe systems of interacting bosons in the mean field approximation
(see
e.g. \cite{Yuri-RevModPhys1989,Kosevich1990,Kaup1990,PerezGarcia1997,Pitaevskii&StringariBook}),
and in the context of fluid mechanics (\cite{NLSfluid}). We now
introduce a formalism which is well known in these domains and will
prove again very useful in the present context of quadratic mean field
games.

\subsection{Ehrenfest's relations and conservation laws}
\label{Sec:Ehrenfest-Conservation}

By analogy with the NLS Equation \cite{Kosevich1990}, we introduce an
operator formalism on some appropriate functional space and derive the
evolution equations for some quantities of interest. To do so, we
first define a position operator $\hat{{\bf X}}=(\hat{ X}_1, \cdots,
\hat{X}_d)$, where $\hat{X}_\nu$ acts as a multiplication by the
$\nu^{\rm th}$ coordinate $x_\nu$.  We also define a momentum operator
${\bf \hat \Pi} \equiv - \mu \sigma^2  \pmb \nabla$. 
    For an arbitrary operator $\hat O $ defined in terms of
${\bf \hat X} $ and ${\bf \hat \Pi}$, we define its average
\begin{equation}
\label{eq:average}
	\langle \hat O \rangle(t)  \equiv  \Pb \hat O \Gk  = 
\int \mathrm{d} \bx \, \Phi(\bx,t)\; \hat O \; \Gamma(\bx,t)  \;,
\end{equation}
where the couple $(\Phi(t),\Gamma(t))$ defines the
state of the system and evolves according to
Eqs~\eqref{eq:NLS_Phi}-\eqref{eq:NLS_Gamma}.  Note that when $\hat O$
depends only on the position: $\hat O= \hat O (\hat{\bf X})$, the
latter average reduces to the usual mean value with respect to the density:
\begin{equation}
\langle \hat O \rangle(t)=  \int \mathrm{d} \bx \, m_t(\bx) O(\bx) \; .
\end{equation}

Differentiating Equation (\ref{eq:average}) with respect to $t$, one gets,
as for the Schr\"odinger equation \cite{cohen}, the time evolution of
the mean value of an  
observable in terms of a commutator:
\begin{equation}
  \frac{d}{dt} \langle \hat O \rangle = 
       \langle \frac{\partial \, \hat O }{\partial t} \rangle 
       -
        \frac{1}{ \mu \sigma^2}\langle
           [\hat O,\hat H]\rangle \; ,
\end{equation}
where we have introduced the Hamiltonian 
\begin{equation} \label{eq:Ham}
\hat H= - \frac{{\bf \hat \Pi}^2}{2 \mu} - \tilde V[m_t]({\bf \hat X})
\; . 
\end{equation}
In particular,  one gets
\begin{align}
\frac{d}{dt}  \langle {\bf \hat X} \rangle  & = 
     \frac{\langle {\bf \hat \Pi} \rangle}{\mu} \; , 
\label{eq:Xdot} \\
\frac{d}{dt} \langle {\bf \hat \Pi} \rangle  &= 
     \langle {\bf \hat F}[m_t] \rangle \;   ,
 \label{eq:Pdot}
\end{align}
where $ {\bf \hat F}[m_t] \equiv - \nabla \tilde
V[m_t]({\bf \hat X})$ is named by analogy the ``force''
operator.    In the same way, introducing  the
variance $\Sigma_\nu$  of the
  $\nu^{\rm th}$ coordinate, $(\nu = 1, \cdots, d)$ , 
\begin{equation} \Sigma_\nu \equiv \sqrt{\langle \hat{X_\nu}^2 \rangle - \langle
\hat X_\nu \rangle^2 }
\end{equation}
and the averaged ``position-momentum'' correlator for the $\nu^{\rm th}$
coordinate
\begin{equation} \label{eq:Lambda}
\Lambda_\nu \equiv \langle \hat X_\nu \hat \Pi_\nu 
+ \hat \Pi_\nu \hat X_\nu \rangle -2 \langle \hat X_\nu  \rangle
\langle \hat \Pi_\nu  \rangle  
\end{equation}
one  has
\begin{align}
     \frac{d}{dt}  \Sigma_\nu  & = \frac{1}{2\mu} \frac{\Lambda_\nu}{\Sigma_\nu }
\; , \label{eq:S2dot} \\
  \frac{d}{dt} \Lambda_\nu & = 2\, \bigl(\langle \hat X_\nu
  \hat{F_\nu}[m_t] \rangle - \langle \hat X_\nu\rangle\langle
  \hat{F_\nu}[m_t] \rangle\bigr) + \frac{2}{\mu} \,\bigl(\langle
  \hat{\Pi_\nu^2}\rangle -\langle \hat{\Pi_\nu^2}\rangle^2\bigr)\;
  . \label{eq:Lambdadot}  
\end{align}

Furthermore, when local interactions are assumed,
\begin{equation}
 \tilde V[m](\bx) = U_0(\bx) +   f[m(\bx)] \; , 
\label{eq:Vlocal}
\end{equation}
 the mean force depends only on the external potential
\begin{align}
\langle \hat F \rangle  & = \langle \hat F_0  \rangle  \label{eq:F} \\
\langle \hat X \hat F  \rangle &
= \langle \hat X F_0 \rangle - \int d\bx \, \bx m_t(\bx) f'[m_t(\bx)]
\label{eq:XF} \; ,
\end{align}
where $ \hat F_0  \equiv -\nabla_\bx U_0(\hat \bX)$ ,
and Eqs.~\eqref{eq:Pdot}-\eqref{eq:Lambdadot} can be simplified
accordingly.

Finally, with such interactions,  we can introduce an action functional
\begin{equation} \label{eq:S}
     S[\Phi,\G]
  \equiv 
 \int_0^T dt  \int_{\mathbb{R}^d } dx
  \left[
   -\frac{\mu \sigma^2}{2}\bigl(  \Phi (\partial_t \Gamma) -  (\partial_t \Phi ) \Gamma  \bigr)  
    - \frac{\mu \sigma^4}{2} \nabla \Phi .\nabla \Gamma 
   + \Phi\,U_0(\bx) \, \Gamma    + F[ \Phi \,\Gamma ] \right] \; ,
\end{equation}
where $F(m)= \int^m f(m') dm'$ , which extremals are solutions of the
system (\ref{eq:NLS_Phi}, \ref{eq:NLS_Gamma}). Indeed, the variational
equation ${\delta S}/{\delta \G} = 0$ (respectively
${\delta S}/{\delta \Phi} = 0 $ ) is equivalent to
Eq.~\eqref{eq:NLS_Phi} (respectively Eq.~\eqref{eq:NLS_Gamma}) with
$\tilde V[m]$ in the form Eq.~\eqref{eq:Vlocal}.  This property will
provide us with a variational approximation scheme for the solutions
of the MFG system. Note however that the boundary conditions
associated with equations \eqref{eq:NLS_Phi} and \eqref{eq:NLS_Gamma}
have to be carefully taken into account.  For a pair $(\Phi,\G)$
solving the MFG equations, the action Eq.~(\ref{eq:S}) can be
rewritten as
\[ S[\Phi,\G] = - \int dt \,d \bx  \left[
    \frac{\mu \sigma^2}{2}\bigl( \Phi(\partial_t \Gamma)  -(\partial_t \Phi ) \Gamma  \bigr)   \right] \; +\; \int dt\,
  E_{\rm tot}(t) \]
where we have introduced the  ``total energy'', 
\begin{equation} \label{eq:Etot}
E_{\rm tot}(t)  \equiv E_{\rm kin} + E_{\rm pot} + E_{\rm int}
\end{equation}
with  ``kinetic'', ``potential'', and ``interaction'' energies
defined respectively as
\begin{align}
E_{\rm kin} &= \frac{1}{2\mu} \langle {\bf \hat \Pi}^2 \rangle,\\ 
E_{\rm pot}  &=  \langle U_0(\hat \bx)  \rangle,\\
E_{\rm int}  &=   \int d\bx F[m(\bx,t)].
\end{align}
The integrand in the action functional \eqref{eq:S} does not depend explicitly on time, so that by Noether theorem, 
there is a conserved quantity along the trajectories. This conserved Noether charge is (due to the sign conventions here) minus the previously defined total energy,
so that 
\begin{equation}
\frac{d E_{\rm tot}(t)}{dt} = 0
\end{equation}
and for any pair  $(\Phi(\bx,t),\G(\bx,t))$ solving the MFG equations, one has 
 \begin{equation}
 S[\Phi, \G] = - \frac{\mu \sigma^2}{2} \int dt \,d \bx\left[ 
  \Phi   (\partial_t \Gamma)  -  (\partial_t \Phi ) \Gamma \right] \; +\;  E_{\rm tot} T  \;  .
    \end{equation}

\subsection{Exactly solvable cases}
\label{sec:solvable}

The list of completely solvable MFG models is up to now rather short, and mainly restricted to stationary settings  \cite{GomesUnpublished}, this situation being most probably due to the difficulties encountered when working  within the original representation
\eqref{eq:hjb}-\eqref{eq:kol}. Hereafter, we add a few examples to that list, by considering first situations in which  either the external potential $U_0(x)$ or the interaction term $V[m]$ 
 is absent or fully negligible.  
These cases will be also used in the following sections as starting
points to develop a perturbative approach when  both terms are present in the potential but one is significantly larger than the other and dominates the optimization
process.

We conclude the present contribution to that list   with the case of an 
  harmonic external potential and local interactions, 
  for which exact solutions can be found for 
  rather specific boundary conditions but no  constraint on
  the relative strength between the two terms of the  potential.

\subsubsection{Non interacting case}
\label{sec:free-case}

We consider first the uncoupled case $V[m] = 0$ so that the potential
reduces to $U_0(\bx)$.  The absence of interactions means that the MFG
is not really a game anymore since the agents  behave independently of
the strategies of the others.  However, this degenerate case appears
naturally in perturbative approaches to weakly interacting regimes
which will be considered in Section~\ref{sec:perturbative}.

Here the potential reduces to the density independent part, $\tilde
V[m](\bx) = U_0(\bx)$  so that  the 
Eqs.~(\ref{eq:NLS_Phi}) and (\ref{eq:NLS_Gamma}) become actually
linear, 
\begin{align}
\mu \sigma^2 \partial_t \Phi & =  + \hat H_0 \Phi,  \qquad \qquad (\Phi(x,T) =
\Phi_T(x))  \label{eq:LS_Phi}\\ 
\mu \sigma^2 \partial_t \G  & =  - \hat H_0 \G,  \qquad \qquad (\G(\bx,0) =
\frac{m_0(\bx)}{\Phi(\bx,0)}) \label{eq:LS_Gamma} 
\end{align}
with $\hat
H_0$ the linear operator
\begin{equation}
H_0= - \frac{{\bf \hat \Pi}^2}{2 \mu} - U_0(\bx)  \; .
\end{equation}

In this case, the solutions of the system
Eqs.~(\ref{eq:LS_Phi})-(\ref{eq:LS_Gamma}) can be expressed in terms
of the eigenvalues $ \lambda_0 \leq \lambda_1\leq
\ldots $ and the associated eigenvectors
$\psi_0(\bx),\psi_1(\bx),\ldots$ of $\hat H_0$.  Explicitly one has
\[ \left\{
  \begin{aligned}
   \Phi(\bx,t) & = \varphi_0 \,e^{-\frac{\lambda_0(T-t)}{\mu
       \sigma^2}}\psi_0(\bx) 
    +   \varphi_1 \,    e^{-\frac{\lambda_1(T-t)}{\mu
       \sigma^2}} \psi_1(\bx) 
    +  \ldots\\ 
    \G(\bx,t) & =  \gamma_0  \,e^{-\frac{\lambda_0 t}{\mu \sigma^2}}
    \psi_0(\bx) 
     +  \gamma_1  \, e^{-\frac{\lambda_1 t}{\mu \sigma^2}}
    \psi_1(\bx) 
     +   \ldots 
  \end{aligned}
\right. \; .
\]
 Assuming $\hat H_0$ non degenerate and the eigenfunctions  $\psi_k$
 normalized,  the coefficients  $\{\varphi_k\}_{k=1,2,\cdots}$ are
 fixed by boundary conditions at $t=T$,
\begin{equation} \label{eq:Phi_k}
\varphi_k = \ \int d\bx \,
\psi_k(\bx) \Phi_T(\bx) \; ,
\end{equation} 
which in particular specifies the expression of $\Phi(\bx,0)$,
\begin{equation}
\Phi(\bx,0)= \varphi_0\, e^{-\frac{\lambda_0T}{\mu
    \sigma^2}}\psi_0(\bx)+ \varphi_1 \, e^{-\frac{\lambda_1T}{\mu
    \sigma^2}}\psi_1(\bx)+\ldots \; .
\end{equation} 
This then fixes the initial value $\G(\bx,0)$, and thus the
coefficients $\{\gamma_k\}_{k=0,1,\cdots}$ of $\G(\bx,t)$ as
\begin{equation} \label{eq:Gamma_k}
\gamma_k =  \int d\bx \,\psi_k(\bx) \,
\frac{m_0(\bx)}{\Phi(x,0)} \; .
\end{equation}

This spectral analysis allows us to see how, in the non-interacting
case, the asymptotic solution converges to the ergodic solution away
from the time boundaries when the horizon $T$ becomes very
large. Indeed, introducing the characteristic convergence  time
\[\tau_{\rm erg} = \mu \sigma^2 / (\lambda_1 -\lambda_0)\]
 one has
\begin{align} \G(\bx,t) &\simeq \gamma_0\, e^{-\frac{\lambda_0 t}{\mu
      \sigma^2}}  \psi_0(\bx)&\mbox{for all } t & \gg \tau_{\rm erg}\\ 
 \Phi(\bx,t) &\simeq \varphi_0 e^{-\frac{\lambda_0(T-t)}{\mu
     \sigma^2}}\psi_0(\bx)& \mbox{for all } t & \ll T- \tau_{\rm erg} 
\; .
 \end{align}
 Hence, when both conditions are fulfilled, the density $m(\bx,t)$
 becomes asymptotically time independent  as,
\[
m(\bx,t) \simeq \gamma_0 \varphi_0\, e^{-\frac{\lambda_0 T}{\sigma^2}}
\Psi_0^2(\bx) \quad  \hbox{ for all } \tau_{\rm erg} \ll t
\ll T-\tau_{\rm erg}. 
\]  
Normalization imposes that $\gamma_0 \varphi_0 \,e^{-\frac{\lambda_0
    T}{\sigma^2}}=1 $, so that in the limit of large
optimization time, the density profile converges exponentially fast
(with the characteristic time $\tau_{\rm erg}$)  to a time independent profile:
\begin{eqnarray}
\lim_{T\to \infty} \|m(\bx,t)-m^e(\bx)\| \le C
e^{\displaystyle{-\frac{t}{\tau_{\rm erg}}}}\\ 
\lim_{T\to \infty} \|m(\bx,T-t)-m^e(\bx)\| \le C e^{\displaystyle{-\frac{t}{\tau_{\rm erg}}}}
\end{eqnarray}
for all time $t$, with $C$ a constant.  Furthermore, the solution of
the ergodic problem Eq.~\eqref{eq:MFG-erg} for a non interacting mean
field game is given by $\lambda^e =\lambda_0$, $u^e(\bx)= - \mu
\sigma^2 \log \psi_0(\bx) + c$ and $m^e(\bx) = \psi_0^2(\bx)$.

Thus any choice of an Hamiltonian $\hat H_0$ with explicitly known
eigenstates would lead to an exactly solvable non-interacting mean
field game problem. The list of analytically diagonalisable $\hat H_0$
indeed contains quite a few systems, among which the case of quadratic
potentials.  Furthermore, for one-dimensional systems, and more
generally for classically integrable Hamiltonian of arbitrary
dimensions, very good approximations can be obtained based on the EBK
approximation scheme \cite{EBK}.  We also note here that the ergodic
problem requires only the knowledge of the eigenstate $\psi_0$
associated with the smallest eigenvalue $\lambda_0$, and the rate of
convergence to it depends only on the first two eigenvalues.

\subsubsection{Local attractive interactions in the absence of
  external  potential.}  
\label{sec:zero-U}

We now turn to the opposite cases when the external potential
$U_0(\bx)$ is negligible with respect to interactions.  More
specifically we consider one dimensional models in which the
interaction term Eq.~\eqref{eq:NLinInt} is local, with the
particular form
\[ 
V[m](x,t) = f[m(x,t)] =  g \, m(x,t)^\alpha  
\] 
with $\alpha > 0$ and $g > 0$. It includes the simple linear form of
the interaction potential Eq.~\eqref{eq:SRLInt} for $\alpha =1$.  In
such cases, the stationary (ergodic) problem Eq.~\eqref{eq:NLSergo}
reduces to a generalized Gross-Pitaevskii equation
\begin{equation} \label{eq:GP}
- \frac{ \mu\sigma^4}{2} \partial^2_{xx} \psi^e 
-  g (\psi^e)^{2 \alpha+1} =  \lambda^e \psi^e \; .
\end{equation}
The lowest energy state can be computed using a known procedure
\cite{Pitaevskii&StringariBook}, that we recall for convenience in
Appendix~\ref{app:GP}. It is associated with an energy
\begin{equation}
\lambda^e 
= -\frac{1}{4} \; \left(
\frac{\Gamma(\frac{2}{\alpha})}{\Gamma(\frac{1}{\alpha})^2}
\right)^\frac{2\alpha}{2-\alpha} \;   
\left( \frac{g}{\alpha+1} \right)^\frac{2}{2-\alpha} \;
\left( \frac{2\alpha}{\mu\sigma^4} \right)^\frac{\alpha}{2-\alpha} 
\end{equation}
( $\Gamma(\cdot)$ is the Euler's Gamma function), and has the following
expression 
\begin{equation} \label{eq:soliton}
  \psi^e(x) = \psi_M\;\left[ \cosh( \frac{x-x_0}{\eta_\alpha}) \right]^{-
  \frac{1}{\alpha}} \; ,
\end{equation}
 where the maximum value $\Psi_M$ reads
\begin{equation}
\psi_M= \left( \frac{\lambda^e (\alpha +1)}{g}\right)^{1/2 \alpha} \; .
\end{equation}
The stationary solution is a localized density around some
arbitrary point $x_0$ (a soliton in the language of the NLS equation),
and its typical spatial extension
\begin{equation} \label{eq:eta}
\eta_\alpha=\frac{2}{\sqrt{\alpha}}\;
\left( \frac{\Gamma(\frac{1}{\alpha})^2}{\Gamma(\frac{2}{\alpha})}
\right)^{\frac{\alpha}{2-\alpha}}\,  
\left( \frac{\alpha+1}{2\alpha}\,\frac{\mu\sigma^4}{g}
\right)^\frac{1}{2-\alpha}  
\end{equation}
depends only on the ratio $({\mu\sigma^4}/{g})$ and results from the
competition between the noise which tends to broaden the distribution
and the attractive interactions.
In the particular case $\alpha=1$, the interaction potential becomes
linear (Eq.~\eqref{eq:SRLInt}) and the above expressions reduce
to:
\begin{align}
\lambda^e|_{\alpha=1}&= -\frac{g^2}{8\mu\sigma^4} \; , \\
\Psi_M|_{\alpha=1}&= \sqrt{\frac{g}{4\mu\sigma^4}} \; , \\
\eta_1&=\frac{2\mu\sigma^4}{g}  \; . \label{eq:eta1}
\end{align} 

Two remarks are in order here.  First, the expressions above are
clearly not well-defined for $\alpha=2$.  As we shall discuss in
section~\ref{sec:more-complicated}, this is related to the fact that
the soliton is unstable for $\alpha >2$.  Moreover, we stress that the
generalized Gross-Pitaevskii equation Eq.~(\ref{eq:GP}) is invariant
under translation and therefore the soliton Eq.~(\ref{eq:soliton}) can
be centered around any point $x_0$ of the real axis.  In presence of a
weak but non zero external potential $U_0$ (section~\ref{sec:strong})
and local interactions as above, it will follows that a very good
approximation for the ergodic state will be a soliton centered at the
maximum $x_{\rm max}$ of $U_0$. We also use this property
  hereafter to derive exact results in the case of an external
  quadratic potential.

\subsubsection{Quadratic external potential}
\label{sec:quadra-U}

 In the rest of this section on exact results, we shall use the formal
connection with the NLS equation (\ref{eq:NLS}) to derive
particular exact solutions of the MFG system
\eqref{eq:NLS_Phi}-\eqref{eq:NLS_Gamma} in dimension one, for a local
interaction potential of the form Eq.~(\ref{eq:NLinInt}) and a
quadratic external potential $U_0(x) = -\frac{1}{2} k x^2$, $k\!>\!0$. 
We thus consider a total potential
\[
\tilde V[m](x) = -\frac{k}{2} x^2 + f(m(x)) \; .
\]
 Following \cite{Pitaevskii&StringariBook}, we
use for $\Phi$ and $\G$ the ansatz 
\begin{align}
\Phi(x,t)&=\exp \left[- \frac{\gamma(t)-x P(t)}{\mu \sigma^2}\right]
\, \psi^e(x -X(t)) \label{eq:exact_Phi} \\ 
\G(x,t)  &=\exp \left[+ \frac{\gamma(t)-x P(t)}{\mu \sigma^2}\right]
\, \psi^e(x -X(t)) \; , 
\label{eq:exact_Gamma}
\end{align}
where $\psi^e(x)$ is the  solution
  of the ergodic equation 
\[
- \frac{\mu \sigma^4}{2} \partial^2_{xx} \psi^e(x) - 
\tilde V[(\psi^e)^2](x)  \psi^e(x)  = \lambda^e \psi^e(x) \; ,
\]
which, for small $k$, is well approximated by the expression on the
r.h.s.\ of Eq.~(\ref{eq:soliton}) (with
  $x_0=0$).  Note that for this ansatz the resulting density is
$m(x,t) = \Phi(x,t) \G(x,t) = \psi_e^2(x-X(t))$, and is thus
independent of $P(t)$ and $\gamma(t)$.

Inserting these expressions into the system
(\ref{eq:NLS_Phi}-\ref{eq:NLS_Gamma}), we get the necessary and
sufficient conditions for 
Eqs.~\eqref{eq:exact_Phi}-\eqref{eq:exact_Gamma} to be an exact solution
of the time dependent problem:
\begin{align}
	\dot{P}(t)&=k X(t)  \label{eq:class1-dotP} \\ 
	\dot{X}(t)&=\frac{P(t)}{\mu} \label{eq:class1-dotX} \\
	\dot{\gamma}(t)&=\frac{k}{2}X(t)^2+\frac{P(t)^2}{2 \mu} - \lambda^e \; . \label{eq:class1-dotg} 
\end{align}
The two first equations
describe the motion of the centre of mass $X(t)$
of the density distribution.
The third one can be integrated in
\begin{equation*}
\gamma(t) = \frac{ X(t) P(t)}{2} -  \lambda^e t +\gamma_0 \; .
\end{equation*}
This solution describes the evolution of a density distribution with
finite spatial extension, that we may call ``soliton'' because it
moves without deformation as a classical particle of mass $\mu$ in an
{\em inverted} quadratic potential $U_0(x) = -\frac{1}{2} k x^2$. It
corresponds, however, to rather specific boundary conditions since the
initial density should be of the form $m(x,t=0)=\psi_e^2(x-x_0)$, and
the function $\Phi_T$ specifying the terminal boundary condition
should be of the form $\Phi_T(x) = K \exp\{ x \,p_T/\mu\sigma^2\}
\psi_e(x-x_T)$ where $p_T$ and $x_T$ are related through the mixed
condition $x_T \cosh(\omega T) -p_T \sqrt{k \mu} \sinh(\omega T)
=x_0$. The two constants $\gamma_0$ and $K$ being unessential, this
family of solutions is fully described by only two parameters,
says $x_0$ and $x_T$.

Assuming that initial and final conditions have been chosen as above,
positions of the center of mass $X(t)$ at initial and final times are
then fixed to $X(0) = x_0$ and $X(T) = x_T$, and for all intermediate
times we get
\[
  X(t) = x_0 \; \frac{\sinh(\omega (T-t))}{\sinh (\omega T)}  +
  x_T \; \frac{\sinh(\omega t)}{\sinh (\omega T)} 
\]
with $\omega \equiv \sqrt{k/\mu}$.  In the long horizon limit
$T\to\infty$, apart from initial and final time intervals of order
$\tau_{\rm erg} = 1/\omega$, the center of mass remains localized in a
close vicinity of the unstable fixed point of the external potential
$U_0$.  This is a general feature that we shall discuss in more
details in the following section.

\section{Strongly attractive short ranged interactions I}
\label{sec:strong}

This section is devoted to simple one dimensional models where the
agents have a strong incentive to coordinate themselves. This first
example of asymptotic regime allows us to make a clear exposition of
the main concepts that can be effectively used, leading to a rather
complete understanding of the behavior of mean field game
equations. In the next two sections, we shall use essentially the
same tools,  addressing somewhat
more intricate settings in the same regime in
section~\ref{sec:more-complicated}, and  considering other
asymptotic regimes in section~\ref{sec:perturbative}.
 
We consider here one dimensional models with interaction potentials
which are local and linear as in Equation \eqref{eq:SRLInt}.  The
total potential is therefore of the form
\[
\tilde V[m](x) = U_0(x) + g \, m(x) \; ,
\]
($g>0$), with a weak external potential $U_0(\bx)$, in a sense
explicited below.  We also assume that the initial distribution of
agents $m_0(x)$ is localized and well described by its mean position
and variance.  We postpone to the next section the discussion on
different interaction potentials or initial conditions.

A characteristic feature which can be easily found in the regime of
strongly attractive short-ranged interactions is that the agents have
a strong incentive to form compact groups evolving coherently which,
by analogy with the NLS nomenclature, we shall call
``solitons''\cite{MAG2008}.  The initial and final boundary conditions
will eventually be an obstruction for the existence of these solitons
for a short period of time $\tau^*$ close to $t=0$ and $t=T$, with
$\tau^* \to 0$ in the limit when the interaction strength $g \to
\infty$.  However, for a sufficiently large time horizon $T$, we
expect that the dynamics of such solitons dominates for a large time
interval of order $[\tau^*,T-\tau^*]$. This naturally raises a few
questions that split in two sets: On the one hand, we have to
understand what are the shape and characteristic scales of these
solitons, and how and how fast they form near $t=0$ and disappear near
$t=T$.  On the other hand, and maybe more importantly since it
dominates most of the time interval, we need to understand what govern
their dynamics.  We first address this simpler question on the
dynamics of the solitons, and will consider in a second stage their
formation and destruction near the time boundaries.

\subsection{Dynamics of the solitons}
\label{sec:SolitonDynamics}

In the limit of large interaction strength $g \to \infty$, and
excluding a neighborhood of time boundaries, we can assume a strongly
localized density of agent $m(x,t)$ with a short characteristic length
$\eta$.  Indeed, for a strength $g$ large enough, the variations of
$U_0$ on the scale $\eta$ can be considered as weak, and in particular
the variations of the external potential around any point, $\delta U_0
= \nabla U_0 .\delta \bx + \sum_{\gamma,\gamma'}
(\partial^2_{\gamma,\gamma'} U_0) \delta x_\gamma \delta x_{\gamma'} +
\cdots$, are dominated by the first term $\nabla U_0 .\delta \bx $ for
a displacement of order $|\delta \bx| \sim \eta$.  In that case,
denoting $\bX_t = \langle \hat \bX \rangle (t)$ the average position
of the soliton, and $P_t = \langle \hat \Pi \rangle (t)$ its average
momentum, the Ehrenfest relations Eqs.~(\ref{eq:Xdot}) and
(\ref{eq:Pdot}) together with Eq.~(\ref{eq:F}) reduce to
\begin{align}
\frac{d}{dt} \bX_t  & =  \frac{\bP_t}{\mu} \; ,  \label{eq:XdotP}\\
\frac{d}{dt} \bP_t  &= \langle F_0 (\bX_t) \rangle \simeq
-\nabla U_0 (\bX_t) \;
. \label{eq:PdotP} 
\end{align}
We again recognize the classical dynamics of a point particle of mass
$\mu$ evolving in the potential $U_0(\bx)$ as in the particular
example of the quadratic external potential studied at the end of
previous section.

However, unlike classical mechanics, where a given trajectory is fully
specified by its initial position and momentum, a mean field game
problem is defined through mixed initial and terminal conditions.  In
the present setting, initial formation and final destruction of a
soliton should occur fast enough that neither the position of the
density center of mass nor the mean momentum are expected to evolve in
any significant way in the meanwhile.  We  are thus led to the
following identifications
\begin{align}
  \bX_{t \smeq 0} &= \int d \bx \, \bx \,  m_0(\bx) \label{eq:CI} \\ 
  \bP_{t \smeq T} &= \langle {\bf \hat \Pi} \rangle (T)  
= \mu \sigma^2 \int d\bx \ \left(\nabla \Phi(\bx,T) \right) \G(\bx,T)
 \label{eq:CT1} 
\end{align}
Eq.~(\ref{eq:CI}) fixes the initial position of the trajectory;
Eq.~\eqref{eq:CT1} can be written as 
\begin{equation} 
 \bP_T = - \int d\bx \, m_T(\bx) \nabla u_T(\bx) 
\end{equation}
which, using the final boundary condition in Eq.~\eqref{eq:hjb}, gives
\begin{equation} \label{eq:CT2}
	\bP_T = -\langle \nabla c_T(\bx)\rangle 
        \simeq - \nabla c_T(\bX_T)  
\end{equation}
where the last approximation holds if $m(\bx,T)$ is localized on the
scale of variations of $c_T(\bx)$, which has to be checked afterwards
for consistency.

The dynamics of the soliton is thus the classical dynamics of a point
particle of mass $\mu$ evolving in the potential $U_0(\bx)$, with an
initial condition Eq.~\eqref{eq:CI} for the position at $t=0$ and a mixed
terminal condition Eq.~\eqref{eq:CT2} involving position and momentum at
$t=T$.

It should be stressed however that, compared to the classical
situation for which the initial position and momentum are specified,
such boundary conditions change drastically the qualitative behavior
of the system under study.  To start with, while the specification of
both initial position and momentum entirely determines a trajectory, a
finite number of trajectories may fulfill the mixed conditions
Eqs.~\eqref{eq:CI}-\eqref{eq:CT2}.  One may therefore have to evaluate
the cost functional Eq.~\eqref{eq:cost3} on each of them to select the
correct solution of the MFG problem.  Furthermore, such a mode of
selection indicates that a MFG system may switch abruptly from one
type of trajectory to another under a small variation of some
parameter and possibly of the optimization time, which would
correspond to a genuine {\em phase transition} in the MFG behavior.

The mixed initial-terminal character of the boundary conditions have
also an implication in the context of the ergodic problem
Eq.~(\ref{eq:MFG-erg}) studied in \cite{Cardaliaguet2013}, and to its
relationship with unstable fixed points of the dynamics. This is
presumably a very general feature of the MFG behavior in the limit of
large optimization times $T \to \infty$, and we consider this question
in some details here since soliton dynamics is the simplest setting in
which it appears.

The dynamics described by Eqs.~(\ref{eq:XdotP}) and (\ref{eq:PdotP})
is illustrated in Figure~\ref{fig:soliton-dynamics} for a one
dimensional MFG system for an external potential with a single
maximum.

\begin{figure}[hbt]
\includegraphics[width=7cm,clip]{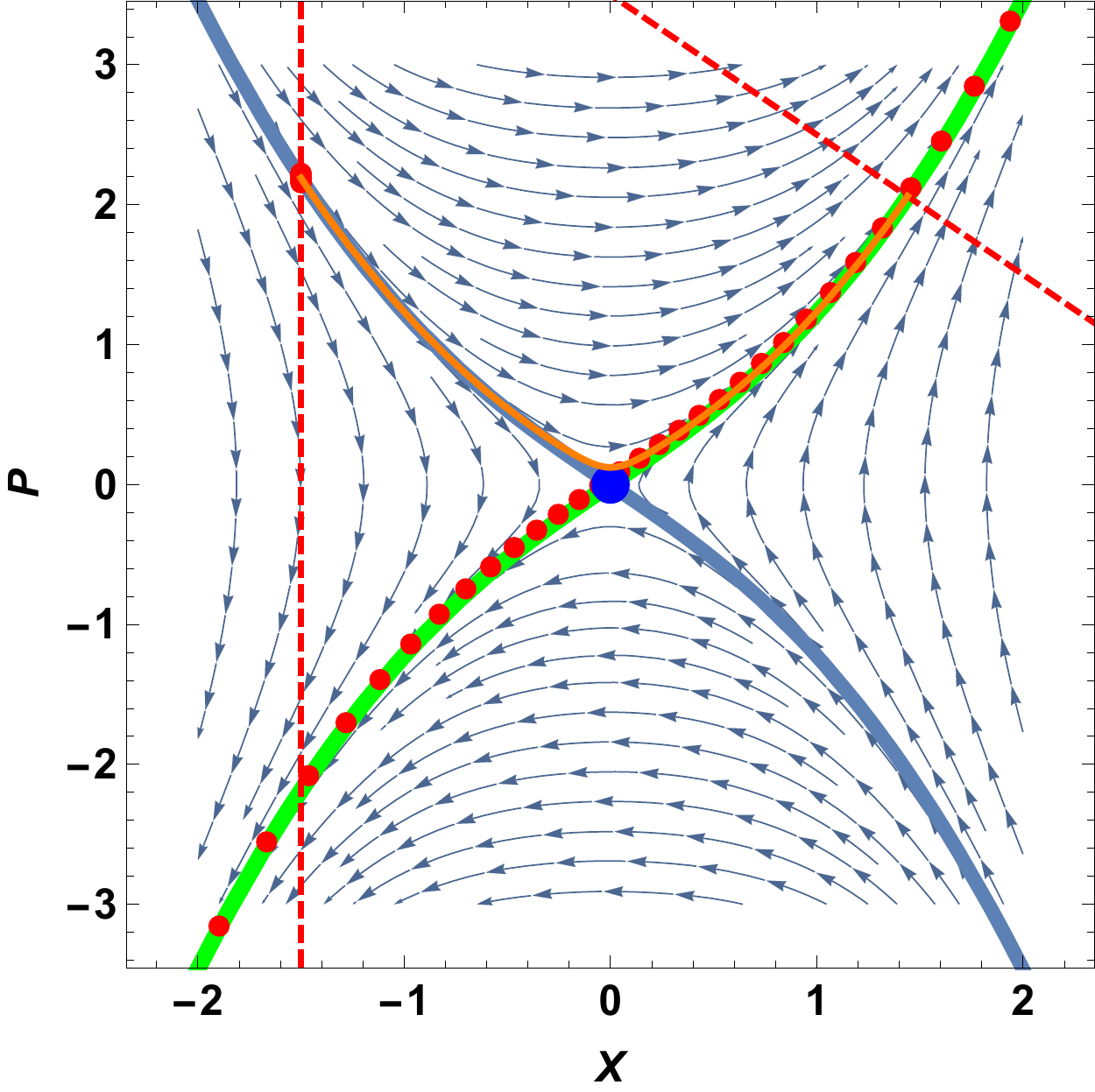}
\caption{ Phase portrait in the plane $(X,P)$ for the
    dynamics of the center of mass of a one dimensional MFG soliton in
    an external potential with a single maximum. The vertical dashed
    red line is the loci of the initial states compatible with initial
    position $X_0=-3/2$; The slanted dashed line is the loci of final
    states compatible with the mixed final condition
    $P_T+X_T=7/2$. The evolution time $T$ determines the actual
    trajectory. For large values of $T$, the soliton has to closely
    follow the stable manifold (blue curve) up to a small neighborhood
    of the unstable fixed point at $(0,0)$ and switch to the unstable
    one (green line) up to its final position.  Here $U_0(x) = -
    \frac{1}{2} x^2 - \frac{1}{4} x^4$, $\mu=1$, $X^0=-3/2$, $c_T(x)=
    x (x-7)/2$ }
\label{fig:soliton-dynamics}
\end{figure}

Let $\tau_{\rm erg}$ the inverse of the Lyapunov exponent associated
with the unstable fixed point $\{X\!=\!X_{\rm max},P\!=\!0\}$ of the dynamics
($X_{\rm max}$ is defined as the
position of the maximum of the potential $U_0$ and is equal to $0$ in
this particular example). For large enough 
values of the time horizon, $T \gg \tau_{\rm erg} $, the MFG system
has to spend most of the time in a neighborhood of the fixed point,
and the dynamics between initial and final
conditions is dominated by the associated stable and unstable
manifolds, respectively,
 \begin{align} 
W^s &= \{(X(0),P(0)) \in \mathbb{R}  \mbox{ such that }
 (X(t),P(t)) \to (X_{\rm max},0) \mbox{ as } t \to +\infty \} \\
 W^u &= \{ (X(0),P(0)) \in \mathbb{R}  \mbox{ such that }
(X(t),P(t)) \to (X_{\rm max},0) \mbox{ as } t \to -\infty \} \; .
\end{align} 
In fact, trajectories are essentially identical for all $T \gg
\tau_{\rm erg}$: They start from the intersection of the stable
manifold $W^s$ with the line of initial condition $X = X_{0}$, closely
follow $W^s$ until they reach a very small neighborhood of the fixed
point $(X,P)= (X_{\rm max},0)$ and then switch to the unstable
manifold $W^u$ that they follow until they reach the intersection of
$W^u$ with the line $P = - \nabla C_T(X)$ specifying the terminal
conditions.  Of course for large but finite $T$ the actual
trajectories are slightly off $W^s$ and $W^u$ and the larger $T$ the
closer to these manifolds they are, and thus the longer they stay in
the immediate neighborhood of $(X_{\rm max},P\!=\!0)$.  \sout{This
  simple situation may be more complex in the presence of more than
  one unstable fixed point (corresponding to more than one local
  maxima of $U_0$).}  Moreover, the dynamics before and after the
dominant fixed point essentially decouple: the dynamics on the stable
manifold toward the unstable fixed point is barely affected by a
change in the terminal condition at $t=T$.

For
$U_0$ with a single maxima, the soliton at rest centered on this
maxima can be identified with the ergodic state defined in
\cite{Cardaliaguet2013}.  When the external potential has multiple
local maxima, only the absolute one is associated with the genuine
ergodic state, but solitons localized on the other local maxima may
play the role of ``effective ergodic states'' in some configurations.

\subsection{Initial and final stages: formation  and
    destruction of the soliton} 
\label{sec:soliton-formation}

We turn now to the slightly more delicate question of describing the
formation of the soliton near $t=0$ and its destruction near $t=T$.
We limit ourselves here to the simpler case where the initial density
$m_0(\bx)$ and the final cost function $\Phi_T(\bx)$ can reasonably be
described through a Gaussian ansatz, and postpone to
section~\ref{sec:more-complicated} the case of more general
configurations.

\subsubsection{Variational method  for the Gaussian ansatz}
\label{sec:variational-1}

One very effective approximation scheme for the Non-Linear
Schr\"odinger equation is the variational method
\cite{PerezGarcia1997}.  A valid approach to it is the use of the
action functional Eq.~\eqref{eq:S}, from which the system
\eqref{eq:NLS_Phi}-\eqref{eq:NLS_Gamma} can be derived.  Variational
approximations then amount to minimizing the action only within a
small subclass of functions. Assuming that the agents' density is
going to contract rapidly around its mean value and then move as a
whole toward the optimal position, we  consider the
following ansatz which is a generalization
  of the exact solution \eqref{eq:exact_Phi}--\eqref{eq:exact_Gamma}
  found for quadratic potential $U_0$:
\begin{align}
\Phi(x,t) & =  \exp \left[ \frac{- \gamma_t + P_t \cdot  x}{\mu\sigma^2}  \right] 
 \frac{1}{\left( 2\pi\Sigma_t^2 \right)^\frac{1}{4}}
\exp \left[-\frac{(x- X_t)^2}{ (2\Sigma_t)^2} 
(1 - \frac{\Lambda_t}{\mu \sigma^2} )\right]\; .
 \label{eq:AnsatzPhi-1d} \\
\G(x,t) &= \exp \left[ \frac{+\gamma_t - P_t \cdot x }{\mu \sigma^2} \right] 
\frac{1}
{\left( 2\pi\Sigma_t^2 \right)^\frac{1}{4}}
\exp \left[-\frac{(x- X_t)^2}{ (2\Sigma_t)^2} 
(1 + \frac{\Lambda_t}{\mu \sigma^2} )\right]  
\; ,   \label{eq:AnsatzGamma-1d} 
\end{align}
Within this ansatz, the resulting density $m(x,t) = \G(x,t) \Phi(x,t)$ reads 
\[
m(x,t) =  \frac{1}
{\sqrt{2\pi \Sigma_t^2}}
\exp \left[-\frac{(x- X_t)^2}{2(\Sigma_t)^2}  \right] \; ,
\]
which is a Gaussian centered in $X_t$ with standard
deviation $\Sigma_t$.  Furthermore, for $\Gamma$ and $\Phi$ given by
Eqs.~(\ref{eq:AnsatzGamma-1d})-(\ref{eq:AnsatzPhi-1d}),
\begin{align}
	\langle {\hat \Pi} \rangle &=  P_t  \; , \\
	\langle {\hat \Lambda} \rangle &= \Lambda_t \; ,
\end{align}
with $ \hat \Lambda$ defined by Eq.~(\ref{eq:Lambda}).  $P_t$ is thus
the average momentum at time $t$, and $\Lambda_t$ the average
position-momentum correlator of the system.

Inserting that variational ansatz into the action Eq.~(\ref{eq:S}), we
get $\displaystyle \tilde S = \int_0^T \tilde L(t) dt$ with the
Lagrangian $\tilde L = \tilde L_\tau + \tilde E_{\rm tot}$ where
\begin{align} \label{eq:ltau}
\tilde L_\tau &= - \int dx \frac{\mu\sigma^2}{2}\left[  \Phi(x,t)  (\partial_t \Gamma(x,t))- 
  (\partial_t \Phi(x,t)) \Gamma(x,t) \right]
\\ 
& = \dot P_t X_t - \frac{\Lambda_t }{2\Sigma_t}    \dot \Sigma_t -
\dot \gamma_t +\frac{1}{4} \dot\Lambda_t 
\end{align}
with the [conserved] total energy 
\begin{equation}
\tilde E_{\rm tot} \equiv \tilde E_{\rm kin} + \tilde E_{\rm int} +
\langle U_0(x) \rangle \; , \label{eq:Etot-tilde-1d}
\end{equation}
where
\begin{align}
\tilde E_{\rm kin} & =  \frac{P^2_t}{2\mu}
 +\frac{\Lambda_t^2 -\mu^2 \sigma^4 }{8\mu \Sigma_t^2} \; ,
  \label{eq:Ekin-tilde-1d} \\ 
\tilde E_{\rm int}  & = \frac{g}{4\sqrt{\pi}\Sigma_t} \; .
\label{eq:Eint-tilde-1d} \\
 \langle U_0(x) \rangle & = \int  d\bx \, U_0(x) \, m(x,t) 
\; .
 \label{eq:Epot-tilde-1d} 
\end{align}

From now on, we choose $\gamma_t = ({\Lambda_t}/{4})$ 
so that the last two terms in the last line of Eq.~\eqref{eq:ltau}
cancel.  Minimizing the reduced action functional with respect to a
variation of the parameters gives first the evolution equations for
$X_t$ and $P_t$:
\begin{align}
	\dot X_t &=\frac{P_t}{\mu}  \label{eq:Xdot2-1d}\\
	\dot P_t &= -
\int  d\bx \, \nabla U_0(x) \, m(x,t) \, = - \langle \nabla U_0
\rangle_t \; ,
                                       \label{eq:Pdot2-1d}
\end{align}
which have the same form as Eqs.~(\ref{eq:Xdot})-(\ref{eq:Pdot}).  The
r.h.s.\ of Eq.~(\ref{eq:Pdot2-1d}) depends both on $X$ and $\Sigma$,
and generally couples the motion of the center of mass to the shape of
the distribution.  However, as soon as $m(x,t)$ is sufficiently narrow
with respect to the scale given by the inverse curvature of the
potential $U_0$, the approximation
\sout{Eqs.~(\ref{eq:Xdot2-1d})-(\ref{eq:Pdot2-1d}) reduce to
  Eqs.~(\ref{eq:XdotP})-(\ref{eq:PdotP})} $\langle \nabla U_0
\rangle_t \simeq \nabla U_0(X_t)$ holds, and the dynamics for the
center of mass $(X_t,P_t)$ decouples from the dynamics of
$(\Sigma_t, \Lambda_t)$.  In the rest of this section, we
consider a situation where that the distribution $m(x,t)$ is
sufficiently narrow at all time so that such a decoupling occurs and
focus on the dynamics of $(\Sigma_t,\Lambda_t)$. In
  particular, the energy of the center of mass, $P_t^2/2\mu + \langle
  U_0(x) \rangle$, is separately conserved and can be dropped from the
  expression for the total energy.  The evolution equations for the
  reduced system $(\Sigma_t,\Lambda_t)$, are thus:
\begin{align}
        \dot\Sigma_t &=
        \frac{\Lambda_t}{2\mu\Sigma_t} 
                                        \label{eq:Sigma-dot-1d} \\
	\dot \Lambda_t  &=
          \frac{\Lambda_t^2- \mu^2 \sigma^4}{2 \mu\Sigma_t^2}
          +
        \frac{g}{2\sqrt{\pi}\Sigma_t}
        \; .                              \label{eq:Lambda-dot-1d}
\end{align}
These equations have a single stationary state $(\Sigma_*,\Lambda_*)$
with
\begin{align} 
\Sigma_* & =    \sqrt{\pi} \, \frac{\mu\sigma^4}{g} \; , \label{eq:fp-11b}\\
\Lambda_* & = 0 \label{eq:fp-11a} \; .
\end{align}
 The stationary standard deviation $\Sigma_*$ is thus, up to a
 numerical constant of order one, equal to the width of the exact
 soliton solution in the absence of an external potential
 \eqref{eq:eta1}.  The total energy at the fixed point is 
\begin{equation} \label{eq:Etotstar-1d}
\tilde E^*_{\rm tot} =
\frac{1}{8\sqrt{\pi}} \frac{g}{\Sigma_*} 
\end{equation}
(note that $\tilde E^*_{\rm tot} =\frac{1}{2} \tilde
    E^*_{\rm int} = -\tilde E^*_{\rm kin} $).

\subsubsection{Time evolution of the  reduced system 
\texorpdfstring{$(\mathbf{\Sigma}_t,\mathbf{\Lambda}_t)$}{(Sigma(t),Lambda(t))} in the long horizon limit}  
\label{sec:TimeEvolSigmaLambda}

Here again, $(\Sigma_*,\Lambda_*)$ is an unstable fixed point for the
dynamics, so that, in the long horizon limit, all trajectories will
follow the associated stable and unstable manifolds, which fixes the
value of the total energy to $\tilde E_{\rm tot}^*$.  Using
Eq.~\eqref{eq:Sigma-dot-1d} in the expression for the total energy
gives then an autonomous equation for the evolution of $\Sigma_t$ on
the stable or unstable manifolds
\begin{equation}
\frac{\mu}{2}\dot\Sigma_t^2 -\frac{\mu\sigma^4}{8 \Sigma_t^2} +
\frac{g}{4\sqrt{\pi} \Sigma_t}= \tilde E^*_{\rm tot} \; ,
\end{equation}
which can be set in the form
\begin{equation}\label{eq:dSigma2-1d}
  \frac{\dot\Sigma_t}{\Sigma_*} = \mp
  \frac{1}{\tau^*}\left( 1-\frac{\Sigma_*}{\Sigma_t} \right) \; ,
\end{equation}
where the minus (respectively plus) sign in front of the r.h.s.\
describes the stable (respectively unstable) manifold.  The factor
$\tau^*$ is the characteristic time
\begin{equation}\label{eq:tau*-1d}
\tau^*= \sqrt{\frac{\mu\Sigma_*^2}{2 \tilde E^*_{\rm tot}}}=
\sqrt{4\pi}  \, \frac {\Sigma_*}{v_g} \; ,
\end{equation}
with 
\begin{equation}\label{eq:vg}
v_g \equiv \frac{g}{\mu \sigma^2} 
\end{equation}
the characteristic velocity associated with the interactions.

Let's us consider first the formation of the soliton. The initial
distribution $m_0(x)$ fixes the initial standard deviation $\Sigma_0 =
\int dx \, (x-\langle x \rangle)^2 m_0(x)$. In terms of the reduced
variable $q_t= \Sigma_t/\Sigma_*$, Eq.~\eqref{eq:dSigma2-1d} can
then be integrated as
\begin{equation} \label{eq:F-of-z-1d}
	F(q_0)-F(q_t)=\frac{t}{\tau^*}  \; ,
\end{equation}
with $F(q) = + q + \log|1-q|$.

Eq.~(\ref{eq:F-of-z-1d}) is an exact implicit solution for the motion
along the stable manifold of $(\Sigma_*,\Lambda_*)$ towards the fixed
point.  It is interesting however to
consider various limiting regimes which are derived straightforwardly
from the limiting behavior of the function $F(q)$:
\begin{align}
F(q) &\simeq  - q^2       & \mbox{for } q &\ll 1 \; , \\
F(q) &\simeq + \log |1-q| & \mbox{for } q &\simeq 1 \; , \\
F(q) &\simeq + q          & \mbox{for } q &\gg 1 \; . 
\end{align}
We have thus the three following possible behavior depending on the
width of the initial distribution:
\begin{itemize}
\item If the initial distribution of agent is much narrower than the
  length scale $\Sigma_*$ characterizing the soliton, the variance
  $\Sigma^2_t$ increases linearly at a rate $\Sigma^2_*/\tau^*$
   until time $\tau^*$ after which it converges
  exponentially  to $\Sigma_*$ (with the
  characteristic time $\tau^*$).
\item If the initial distribution of agent is already close to
  $\Sigma_*$, it converges exponentially  to $\Sigma_*$
  with the characteristic time $\tau^*$.
\item If the initial distribution of agent is much wider than the
  length scale $\Sigma_*$ characterizing the soliton, the standard
  deviation $\Sigma_t$ decreases linearly at rate $\Sigma_*/\tau^*$ up
  to a time $t_c = (\Sigma_0/\Sigma_*)\tau^*$ from
which is converges exponentially  to $\Sigma_*$
(with the characteristic time $\tau^*$).
\end{itemize}

Considering now the destruction of the soliton near $T$, and again
foreseeing that the density will remain localized on a scale
$\Sigma_T$ that can be assumed small, we can write out a simplified
form of the terminal condition for $\Sigma^2$.  Indeed, the ansatz
Eq.~\eqref{eq:AnsatzPhi-1d} for $\Phi$ implies that $u(x,T) = - \mu
\sigma^2 \log \Phi(x,T)$, and thus reads
\[
u(x,T) = \gamma_T +  \frac{1}{4} \mu \sigma^2 \log(2\pi\Sigma_T^2)  -
P_T x + \frac{\mu\sigma^2 - \Lambda_T}{4 
  \Sigma^2_T} (x-X_T)^2 +\cdots
\]
Assuming the final density localized, we can make a Taylor  expansion for
the terminal condition $u(x,T) = c_T(x)$ near $X_T$
\[
u(x,T) \simeq c_T(X_T) + \frac{d c_T}{dx}\big|_{X_T}\cdot (x-X_T) +
\frac{d^2 c_T}{dx^2}\big|_{X_T}\cdot \frac{(x-X_T)^2}{2}  +\cdots
\qquad .
\]
Identifying term by term the coefficients of  both expansions,  
we recover from the first order term  the terminal
condition Eq.~\eqref{eq:CT2} for the center of mass motion, while the second
order term  gives
\begin{equation}\label{eq:TC-zT-2a}
\frac{d^2 c_T}{dx^2}(X_T) =
\frac{\mu\sigma^2 - \Lambda_T}{2 \Sigma^2_T} 
= \mu \left( \frac{\sigma^2}{2\Sigma_T^2}  - \frac{\dot\Sigma_T}{ \Sigma_T}\Bigr)
=\frac{\mu}{\tau^*}\frac{\Sigma_*}{\Sigma_T}\Bigl(2\frac{\Sigma_*}{\Sigma_T}
-1 \right) \; ,
\end{equation}
where we have used both the evolution equation \eqref{eq:dSigma2-1d}
along the unstable manifold and the relation $\tau^*=2
\Sigma_*^2/\sigma^2$. 

In the reduced notations, $q_T=\Sigma_T/\Sigma_*$, the equation
for the terminal width reads
\begin{equation}
\frac{ q_T-2}{q_T^2}  =\frac{\tau^*}{\mu} \frac{d^2 c_T}{dx^2}(X_T) \; .
\label{eq:TC-zT-2b}
\end{equation}
In the strong interaction limit, $g\to \infty$, the characteristic
time $\tau^*$ goes to zero and the right hand side of this equation
(which is the only term depending on the terminal cost $c_T(x)$)
becomes negligible.  Thus, in accordance with the approximation scheme
used here, one has to take $q_T \simeq 2$, so that the final
distribution $m(x,T)$ has an extension of order $2 \Sigma_*$ and thus
small, as anticipated.  Accordingly, the variance at large times 
is given  by
\begin{equation}
F(q_t)=F(q_T) - \frac{T-t}{\tau^*} \; ,
 \end{equation}
 where $q_T$ is the smallest solution of Eq.~\eqref{eq:TC-zT-2b},
 and $q_t$, $\tau^*$ and $F(q)$ are defined as above.

\subsection{Discussion}

In this section, we have considered the strong positive interaction
regime under three simplifying assumptions : i)  the state space
  is  one dimensional; ii) the interaction between the agents is local
  and linear ($V[m](x) = g m(x)$); and iii) the initial distribution of agents
$m_0(x)$ is reasonably well described by a Gaussian.

Under these hypothesis, the image that emerges for the evolution of
the agent in the state space is extremely simple.  The dynamics is
divided in three stages : the initial formation of the soliton near
$t \! = \! 0$, its propagation, and its final destruction near $t \! =
\! T$.

Actually, for most of the time interval $]0,T[$ the density of agents
$m(x,t)$ is well approximated by a soliton of extension $\Sigma_* \sim
\eta^{(1)}$ (Eq.~\eqref{eq:eta1}), which is the shortest length scale
of the problem.  This soliton evolves as a classical particle in a
potential $U_0(x)$ (i.e. following the Hamilton equations
Eqs.~(\ref{eq:XdotP})-(\ref{eq:PdotP})), with the initial and terminal
conditions Eqs.~(\ref{eq:CI})-(\ref{eq:CT1}).  In the long horizon
limit $T \to \infty$, this motion is furthermore dominated by the
maxima of $U_0(x)$ which correspond to an unstable fixed point of the
dynamics where the ``soliton'' spends most of its time
\cite{Cardaliaguet2013}, while the initial (resp.\ final) motion
takes place along the related stable (resp.\ unstable) manifold.
There may also exists intermediate regimes for the long horizon limit
where other unstable fixed points, when they exist, may also show up
in the dynamics and play the role of ``effective ergodic state''.

This propagation phase is flanked by significantly shorter initial and
final phases where the soliton is respectively formed and
destroyed. When the initial distribution of agent $m_0(x)$ has an
extension $\Sigma_0$ of the order of the stationary value $\Sigma_*$
or smaller, a soliton forms within a typical time of order $\tau^*$
(Eq.~\eqref{eq:tau*-1d}).  For
initial extensions $\Sigma_0$ much larger than $\Sigma_*$, the time of
formation of the soliton is larger, $(\Sigma_0/\Sigma_*) \tau^*$.  In
the final phase, the typical extension of the distribution is of order
$2\Sigma_0$ for strong interactions and the soliton always disappears
in a time of order $\tau^*$.

\section{Strongly attractive short ranged interactions II}
  
\label{sec:more-complicated}

In this section, we continue with the study of the strong positive
coordination regime, but we relax some of the simplifying
assumptions made in section~\ref{sec:strong} concerning the
dimensionality of the space, the form of the interactions, and the
shape of the initial distribution of agents.

 We will show that the  dominant phase of the  dynamics, namely the
propagation of the soliton, is essentially unaffected (or only
trivially affected) by these modifications.   Most of our
discussions here will concern the formation and destruction phases of
the soliton (and in practice we will essentially focus on the former).

This section will be divided in three parts.  In the first one, we
shall extend the variational approach of section~\ref{sec:strong} to
higher dimensionality problems and to different form of the
interaction between agents.  In a second subsection, we shall discuss
the ``collapse'' of the distribution of agents which
does occur for nonlinear local interaction or in higher dimension even
with linear interactions.  Finally, in the last subsection, we shall
discuss the formation of the soliton when the initial distribution
$m_0(x)$ has some structure and cannot be just described by its mean
and variance.

\subsection{Gaussian ansatz in  higher
  dimensions and nonlinear interactions}

In this subsection, we generalize the variational approach of
section~\ref{sec:variational-1} to the case where the ``state space''
of the agents is of dimension higher than one and for non-linear
interaction between the agents of the form $V[m](\bx) =  g
m^\alpha(\bx)$, so that
\begin{equation}\label{eq:NLinPot}
\tilde V[m](\bx) = U_0(\bx) + g \left[m(x)\right]^\alpha \; ,
\end{equation}
with, as  in the previous section, $g>0$, $\alpha>0$ and $U_0(\bx)$
assumed non zero but weak.

We first generalize the variational ansatz
Eq.~(\ref{eq:AnsatzGamma-1d})-(\ref{eq:AnsatzPhi-1d}) to 
\begin{align}
\Phi(\bx,t) & =  \exp \left\{ \frac{-\gamma_t + \bP_t \cdot \bx
   }{\mu\sigma^2}  \right\}
\prod_{\nu=1}^d \left[
\frac{1}{\left(2\pi (\Sigma^\nu_t)^2\right)^{1/4}}
\exp  \left\{-\frac{(x^\nu- X^\nu_t)^2}{ (2\Sigma^\nu_t)^2} 
(1 - \frac{\Lambda^\nu_t}{\mu \sigma^2} )\right\}
\right] 
 \label{eq:AnsatzPhi} \\
\G(\bx,t) &= \exp \left\{\frac{+\gamma_t - \bP_t \cdot \bx}{\mu
    \sigma^2}  \right\}
\prod_{\nu=1}^d \left[
\frac{1}{\left(2\pi (\Sigma^\nu_t)^2\right)^{1/4}}
\exp \left\{-\frac{(x^\nu- X^\nu_t)^2}{ (2\Sigma^\nu_t)^2} 
(1+\frac{\Lambda^\nu_t}{\mu \sigma^2} )\right\}
\right] 
   \label{eq:AnsatzGamma}
\end{align}

Inserting these expressions into
the action Eq.~\eqref{eq:S} (see
appendix~\ref{app:variational}), we get an action functional in
  the variables $(X_t^\nu,P_t^\nu)_{\{\nu=1,\cdots,d\}}$ and
  $(\Sigma_t^\nu,\Lambda_t^\nu)_{\{\nu=1,\cdots,d\}}$.   We consider first the
  equations of motion for the $X_t^\nu$ and $P_t^\nu$:
\begin{align}
	\dot X_t^\nu &=\frac{P_t^\nu}{\mu}  \label{eq:Xdot2}\\
	\dot P_t^\nu &= -  \langle \partial^\nu U_0(\bx) \rangle_t
        \simeq - \partial^\nu U_0(\bX_t)   \label{eq:Pdot2}
 \end{align}
 which decouples from the dynamics of the $\Sigma^\nu$'s and
 $\Lambda^\nu$'s when the approximation in \eqref{eq:Pdot2} is
 assumed. This approach is valid whenever the density of agents is
 sufficiently narrow with respect to the inverse curvature of the
 potential, condition which in the strong positive coordination regime
 will be fulfilled at almost all time (except possibly within a very
 short time near $t=0$ or near $t=T$).  As in the one dimensional
 case, the mean position $\bX$ and mean momentum $\bP$ hence follow
 the motion of a classical particle of mass $\mu$ in the external
 potential $U_0(\bX)$, with initial and terminal conditions which are
 the direct generalization of Eqs.~(\ref{eq:CI})-(\ref{eq:CT2}): 
\begin{align}
X_{t=0}^\nu&=\int d^d\bx x^\nu m_0(\bx)\\
P_{t=T}^\nu&=-  \langle \partial^\nu c_T(\bx) \rangle_T \simeq
- \partial^\nu c_T(\bX_T)  \; .
\end{align}
This motion can be more complex than in the one-$d$ case since the
conservation of the total energy is not sufficient to make the
dynamics integrable anymore. However, in the long horizon limit, it is
still dominated by the maxima of the potential $U_0(\bx_{\rm max})$
and takes place very close to the stable and unstable manifolds of the
unstable fixed point   $(\bX\!=\! \bx_{\rm max}$, $\bP\! =\! 0)$.

We assume from now on that the motion of the center of mass decouples
from the evolution of the shape of the density profile and consider
the dynamics of $(\Sigma_t^\nu,\Lambda_t^\nu)_{\{\nu=1,\cdots,d\}}$
alone. We consider the case of a local interaction of the form $V[m] =
g m^\alpha$ and from the variation of the reduced action, we get
  the following evolution equations:
\begin{align}
  \dot\Sigma^\nu &=
  \frac{\Lambda^\nu}{2\mu\Sigma^\nu }  \; , \label{eq:Sigma-dot} \\
  \dot \Lambda^\nu_t &= 
  \frac{(\Lambda_t^\nu)^2-\mu^2\sigma^4}{2\mu(\Sigma^\nu_t)^2}
  + \frac{2 g\alpha }{\alpha+1} \prod_{{\nu'} =1}^d 
  \left[\frac{1}{\sqrt{\alpha+1} (2\pi)^{\alpha/2}} 
    \left(\frac{1}{\Sigma^{\nu'}_t}\right)^\alpha\right] \; .
\label{eq:Lambda-dot}
\end{align}
These equations admit a single stationary state
$(\mathbf{\Sigma}_*,\mathbf{\Lambda}_*)$, with
\begin{align}
  \Lambda^\nu_* & = 0 \\
  \Sigma^\nu_*  &= \Sigma_* = \left[\frac{4\alpha}{\alpha+1} 
    \left(\frac{1}{(\alpha+1)(2\pi)^\alpha}\right)^{d/2} 
    \frac{g}{\mu\sigma^4}\right]^{-1/(2-\alpha d)} \; .
  \label{eq:Sigma*_d_alphaA}
\end{align}

As in section~\ref{sec:zero-U} for the one dimensional case $d=1$,
there is no solution for $\alpha=2/d$.  We understand here this
critical value of $\alpha$ as the transition between the situation
$0<\alpha<2/d$ where $(\mathbf{\Sigma}_*,\mathbf{\Lambda}_*)$ is a
saddle point for the total energy of the reduced system
Eq.~\eqref{eq:rEtot}, and the situation $\alpha > 2/d$ where it is a
minima, leading to a change of stability.  From a physical point of
view, this means that for $\alpha > 2/d$, attractive interactions
dominate at short distance while diffusion, which tends to disperse
the density $m(x,t)$ dominates at large distance, which makes the
``soliton'' unstable. Note that the stability (resp.\ instability) of
the soliton is associated with instability (resp.\ stability) of
trajectories.

\subsection{Collapse for \texorpdfstring{$d \alpha >2$}{d alpha>2}}  
\newcommand {\qq}{\mathsf{q}}
\newcommand {\pp}{\mathsf{p}}

To get a better picture of the main differences between the regime
$\alpha < 2/d$ where the soliton is stable and the regime $\alpha >
2/d$ where it is unstable, we restrict ourselves to the 1-dimensional
case (thus $\alpha_c = 2$) and introduce the canonical variables
\begin{align}
\qq_t &= \frac{\Sigma_t}{\Sigma_*} \; , \\
\pp_t &= - \frac{\Sigma_*}{2} \,\frac{\Lambda_t}{\Sigma_t} \; ,
\end{align}
The Lagrangian Eq.~(\ref{eq:Ltilde}) reads
\begin{align}
  \tilde L(t) &=  \pp_t  \dot \qq_t  - h(\pp_t ,\qq_t ) \; , \\
  h(\pp ,\qq )  &= - \frac{\pp^2}{2\mu\Sigma_*^2} +
  \frac{\mu\sigma^4}{4\Sigma_*^2} \left[ \frac{1}{2 \qq^2} - 
    \frac{1}{\alpha \qq^\alpha} \right] \; ,
\label{eq:hqqpp}
\end{align}
and the equation of motions takes the canonical form in term of the
Hamiltonian $h(\pp,\qq)$
\begin{align}
  \dot \qq = + \frac{\partial h(\pp ,\qq ) }{\partial \pp} &= -
  \frac{\pp}{\mu\Sigma_*^2} \label{eq:canonical-q} \\
  \dot \pp = - \frac{\partial h(\pp ,\qq ) }{\partial \qq} &= 
  \frac{\mu\sigma^4}{4\Sigma_*^2}\left[\frac{1}{\qq^3}-\frac{1}{\qq^{(\alpha+1)}}
  \right] \; . \label{eq:canonical-p}
\end{align}


With these variables, conservation of the total energy $\tilde E_{\rm
  tot} = - h(\pp,\qq)$ is manifest, and the fact that the Liouville
measure $d\pp\,d\qq$ is conserved (which would be also true for
$d>1$ or in the full problem when variables $(\pp,\qq)$ are coupled
with the global motion $(P,X)$) makes it possible to classify a priory
the fixed points by their stability.

Specifically here, the dynamical system
Eqs.~\eqref{eq:canonical-q}-\eqref{eq:canonical-p} has one fixed point
at  $(\qq_* \smeq 1, \pp_* \smeq 0)$, where the second
derivatives of $h(\qq,\pp)$ are given by
\begin{equation} \label{eq:D2pq}
\frac{\partial^2 h}{\partial^2 \pp }\Big|_{(\qq_*, \pp_*)} =
\frac{-1}{\mu\Sigma_*^2} \; , \qquad
\frac{\partial^2 h}{\partial \pp \partial \qq}\Big|_{(\qq_*, \pp_*)}
= 0 \; , \qquad
\frac{\partial^2 h}{\partial^2 \qq}\Big|_{(\qq_*, \pp_*)}  = 
\frac{\mu\sigma^4}{4\Sigma_*^2} \, \left( 2- \alpha \right) \; .
\end{equation}
We see therefore that, as expected, the stability of the fixed point
is entirely determined by the sign of $(\alpha -2)$.
\begin{itemize}
\item For $ 0< \alpha < 2$, $(\qq^*,\pp^*)$ is a saddle point for
  $h(\qq,\pp)$ and thus an unstable fixed point. The dynamics is in
  that case qualitatively similar to the one of
  section~\ref{sec:TimeEvolSigmaLambda}: Near $t=0$ (formation of the
  soliton), the system starts from the initial $\qq_0$ fixed by the
  initial density $m_0(x)$ and follow the stable manifold of
  $(\qq^*,\pp^*)$, which it approaches exponentially closely on a very
  short time scale.  The destruction of the soliton follows a similar
  scenario, but on the unstable manifold. The typical phase portrait
  in this case is shown in Figure~\ref{fig:portrait}a.
\item If $\alpha > 2$, $(\qq^*,\pp^*)$ is a minima of $h(\qq,\pp)$ and
  thus a stable (elliptic) fixed point for the classical dynamics
  governed by
  Eqs.~\eqref{eq:canonical-q}-\eqref{eq:canonical-p}. \sout{the energy
    value $\tilde E_{\rm tot}$.} For a given set of initial and final conditions, one cannot
    exclude the possibility that a periodic orbits in the neighborhood
    of $(\qq^*,\pp^*)$ turns out to be solution of the equations of
    motion, which would correspond to a kind of breathing of the
    soliton.  Our guess, however, is that these breathing mode, when
    they exist, are not the only solution of the equations of motion,
    and can be eliminated because they do not minimize the cost
    Eq.~(\ref{eq:cost3}) (in the sense that they correspond to local
    minima, but not absolute minima of this cost). In the limit of
  long time horizon, the system will prefer to flow toward other
  fixed points: either a low density
(non-Gaussian) noise-dominated phase (described in our ansatz
  by the limit $q\to \infty$), or a large density phase dominated by
  the interactions (here obtained in the limit $q\to 0$).  This case
  is illustrated in Figures~\ref{fig:portrait}b--\ref{fig:portrait}c
  in the particular case $\alpha=3$. 
\end{itemize}

\begin{figure}[ht] 
\centering
  \begin{tabular}[b]{c}
\includegraphics[width=6cm]{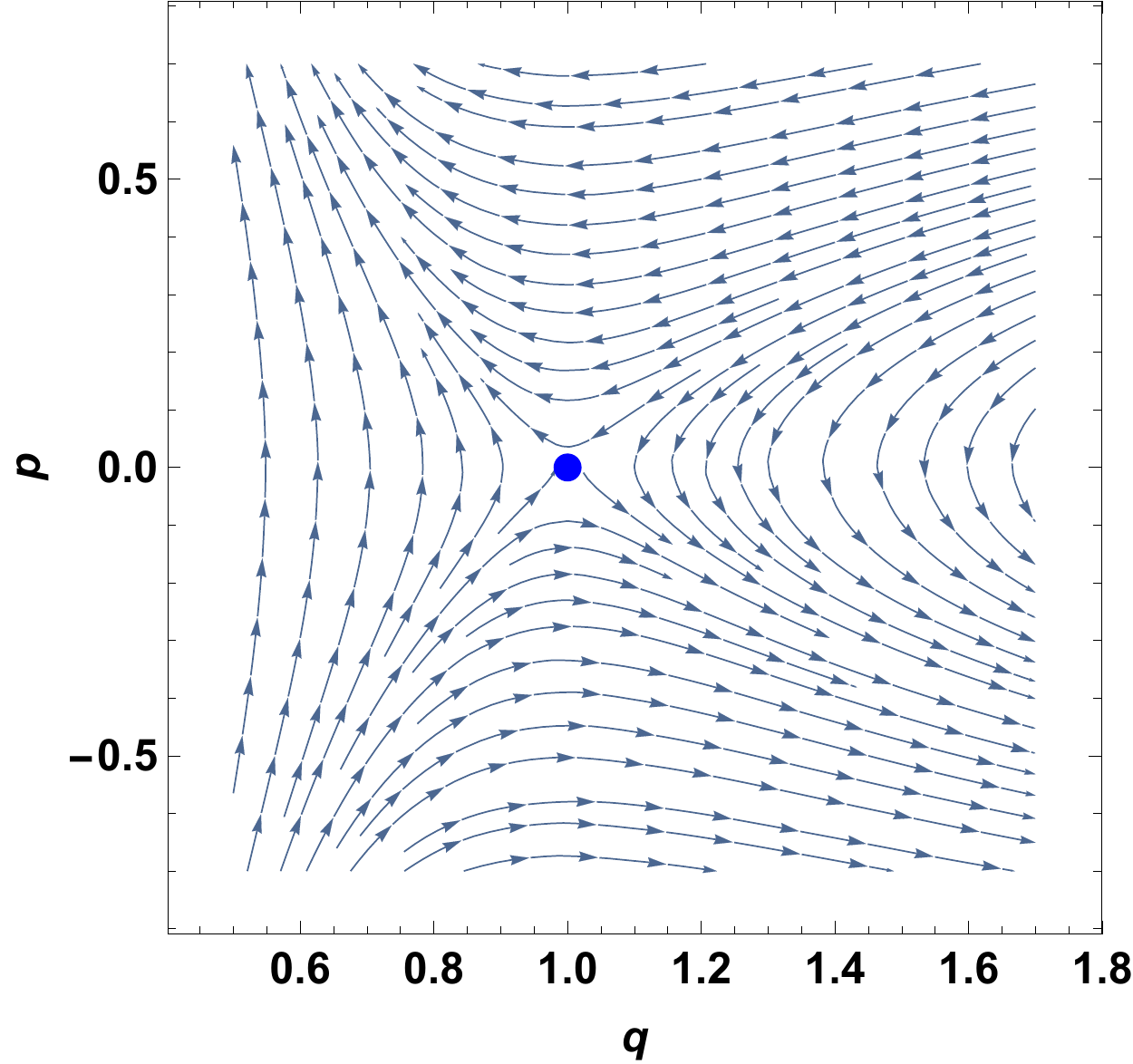}
\\ \small (a)
  \end{tabular} \qquad
  \begin{tabular}[b]{c}
\includegraphics[width=6cm]{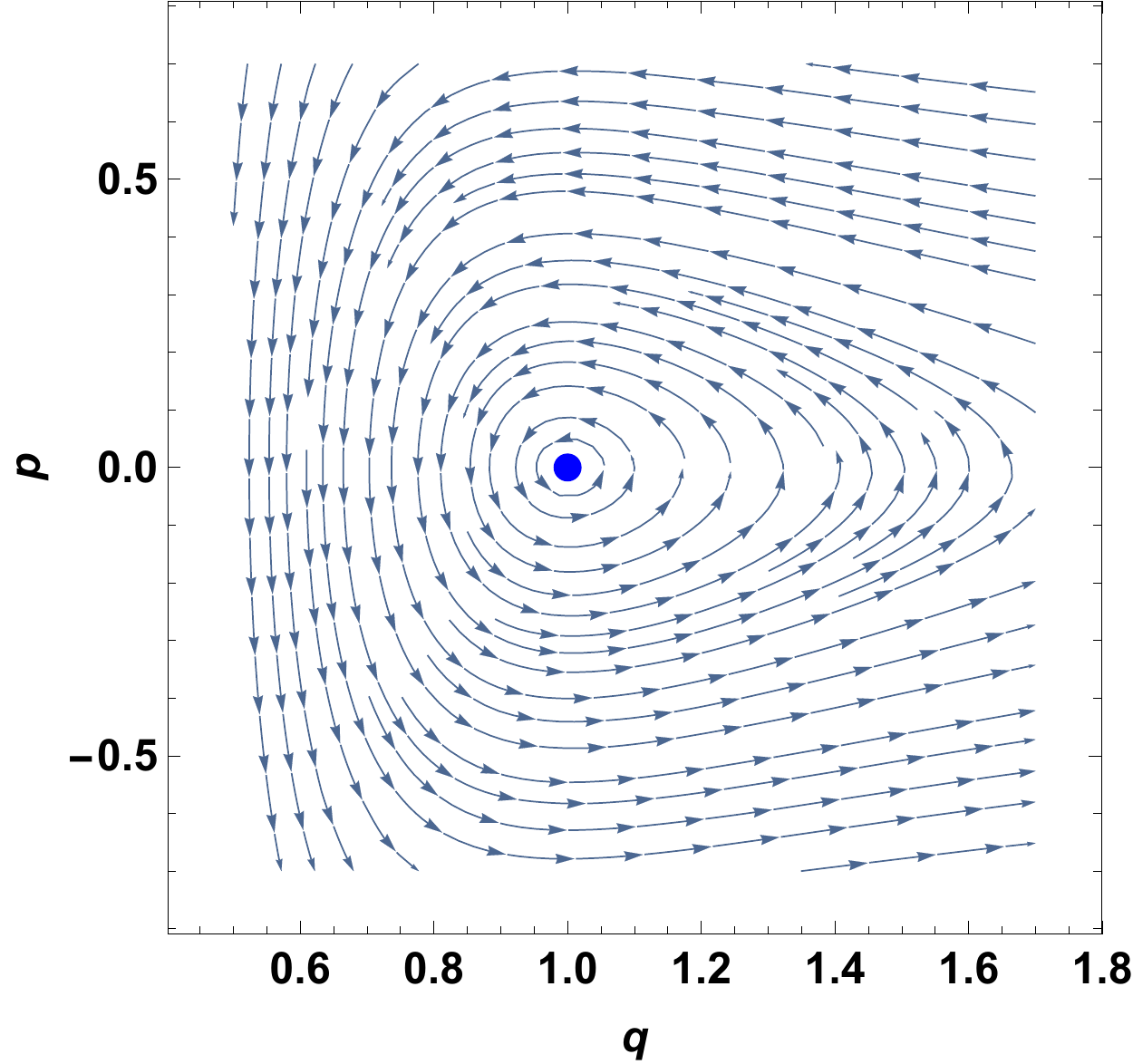}
 \\   \small (b)
  \end{tabular}\\ \vskip.2cm
  \begin{tabular}[b]{c}
  \includegraphics[width=6cm]{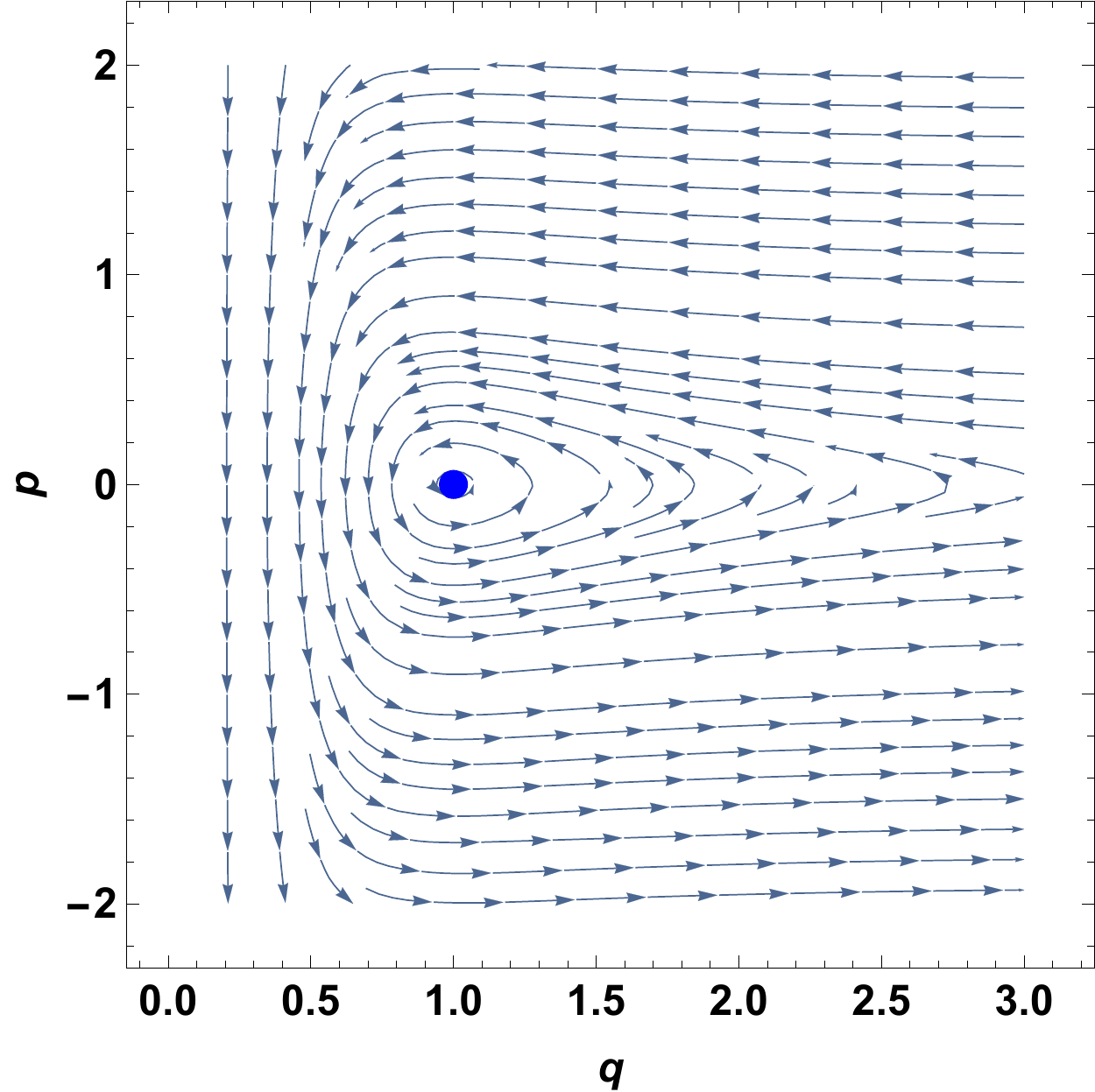}
  \\    \small (c)
  \end{tabular}
  \caption{Phase portrait in the canonical variables $(\qq,\pp)$. (a):
    $0< \alpha< 2$. The fixed point is unstable and long time
    trajectories stay close to the stable and unstable manifold
    ($\alpha=1$, $\mu \sigma^2 =1$ ).  (b) and (c): $\alpha>2$. The
    fixed point is locally stable (b) but non closed trajectories can
    be seen at larger scale (c) ($\alpha=3$, $\mu \sigma^2 =1$).}
\label{fig:portrait}
\end{figure}

Hence, for $\alpha>2$, we have two possible options: either a large
spreading of the distribution, or a collapse.  In the first case,
 namely a large excursion toward large $\qq$'s
(and thus large $\Sigma$'s)  the initial spreading of the density  should
  be large enough so that the noise becomes the dominating
force.  Within the approximation scheme we use here, what the system
does is then to spread out relatively slowly  under the influence of the noise,
and (possibly) re-compactify toward the end (i.e.\ for $t$ near $T$)
if the terminal constraint makes this mandatory.  In practice however,
a system in this configuration is effectively not any more in the
strong interaction regime.  There is no short time scale associated
with the interaction between the agents, and the influence of $U_0(x)$
may become as significant as the one of the noise.  Furthermore since
the distribution of agents  does not remain
localized, the dynamics of $(\qq_t,\pp_t)$, does not
  decouple from the center of mass  $(X_t,P_t)$, and even the validity
of the Gaussian ansatz
Eqs.~(\ref{eq:AnsatzGamma-1d})-(\ref{eq:AnsatzPhi-1d}) becomes
questionable.  The analysis of this regime should actually follow the
line of section~\ref{sec:perturbative}.

Second, if the initial and final conditions select a regime where the
density of agents remains sufficiently large, then the system will
rather choose a large excursion toward $\Sigma_t \to 0$, and thus a
collapse of the density of agents. In that case, we need to consider
explicitly a ``finite-range'' interaction. Indeed, the rational behind
the utilization of a ``zero-range'' interaction potential
Eq.~\eqref{eq:NLinInt} is that the actual range of the interactions is
the smallest length scales in the problem, which cannot hold any more
here since $\Sigma_t$ would eventually become smaller than whatever
this range is.

Let us illustrate this in the case $\alpha=3$  depicted in
Figures~\ref{fig:portrait}b--\ref{fig:portrait}c.  The interaction
$V[m](x) =  g m(x)^3$ can be seen as the $\xi \to 0$ limit of 
\[
V[m](x) = g \int dy_2 dy_3 dy_4 K(x,y_2,y_3,y_4) m(y_2)m(y_3)m(y_4) \;
, \]
with 
\[
K(y_1,y_2,y_3,y_4) \equiv \frac{1}{2 (\sqrt{2 \pi}\xi)^{3}}
\exp\left[-\frac{1}{16 \xi^2} \sum_{i \neq j} (y_i-y_j)^2 \right]
\quad \operatornamewithlimits{\longrightarrow}_{\xi \to 0} \delta(y_1
- y_2)\delta(y_2 - y_3)\delta(y_3 - y_4) \; .
\]
The analysis of this ``finite range'' interaction can be done along
the same line as before, up to the replacement of the interaction
energy term by
\[
\tilde E_{\rm int} = \frac{g}{8} \left( 
  \frac{1}{ \sqrt{2\pi(\xi^2
      +\Sigma_t^2)}}\right)^3
\operatornamewithlimits{\longrightarrow}_{\xi \to 0}\frac{g}{8 (2
  \pi)^{3/2} \Sigma_t^3} 
\] 
(which is indeed the second term of Eq.~\eqref{eq:rEtot} for
$d=1$ and  $\alpha=3$).

With the $(\qq_t,\pp_t)$ variable, the Hamiltonian
Eq.~(\ref{eq:hqqpp}) becomes
\[
   h(p ,q )  = - \frac{p^2}{2\mu\Sigma_*^2}+
   \frac{\mu\sigma^4}{4\Sigma_*^2} \left[ \frac{1}{2 q^2}- 
     \frac{1}{3}\left(1+\frac{\xi^2}{\Sigma_*^2}\right) 
     \left(\frac{\xi^2+\Sigma_*^2}{\xi^2+\Sigma_*^2 q^2} 
     \right)^{3/2}\right]
\]
(which as $\xi \to 0$ indeed correspond to the Hamiltonian
Eq.~\eqref{eq:hqqpp} with $\alpha=3$). 
Here $\Sigma_*$ is the value of $\Sigma$ at a stationary point and is a solution of 
\begin{equation} \label{eq:Sigma*eta}
  \frac{\xi\Sigma_*^4}{(\xi^2+\Sigma_*^2)^{5/2}} = 
  \frac{2 (2\pi)^{3/2}}{3}\frac{\xi\mu\sigma^4}{g} \; .
\end{equation}
Note that the left hand side of Eq.~\eqref{eq:Sigma*eta} depends only
on the ratio $\Sigma_*/\xi$ and has a single maximum at $
\Sigma_*/\xi= 2 $. For $\xi$ (and the right hand side of
\eqref{eq:Sigma*eta}) small enough, Eq.~\eqref{eq:Sigma*eta} has thus
exactly two positive solutions, says, $ \Sigma_*^{(1)}$ and $
\Sigma_*^{(2)}$, which are respectively smaller and larger than
$2\xi$, each one associated to a stationary point for the dynamics;
the second derivative of $H(p,q)$ with respect to $q$ now reads
\begin{equation}
\frac{\partial^2 h}{\partial^2 \qq}\Big|_{(\qq_*, \pp_*)}  
= \frac{\mu\sigma^4}{4\Sigma_*^2} \,
\left(\frac{4\xi^2-\Sigma_*^2}{\xi^2+\Sigma_*^2} \right) 
\end{equation}
while the other two second derivatives remain as in
\eqref{eq:D2pq}. Thus the smallest value $ \Sigma_*^{(1)}$ is
associated with an hyperbolic fixed point and the larger one
$\Sigma_*^{(2)}$ with an elliptic fixed
point.  The corresponding phase portrait is shown on
Figure~\ref{fig:portrait-smooth}, where both fixed points appear.  The
stable ``soliton'' is associated with this new fixed point, and its size
$ \Sigma_*^{(1)}$ is governed by the range of the interaction $\xi$
and not any more by the balance between the strength of the
interaction and the one of the noise.

\begin{figure}[ht] 
\includegraphics[width=6cm]{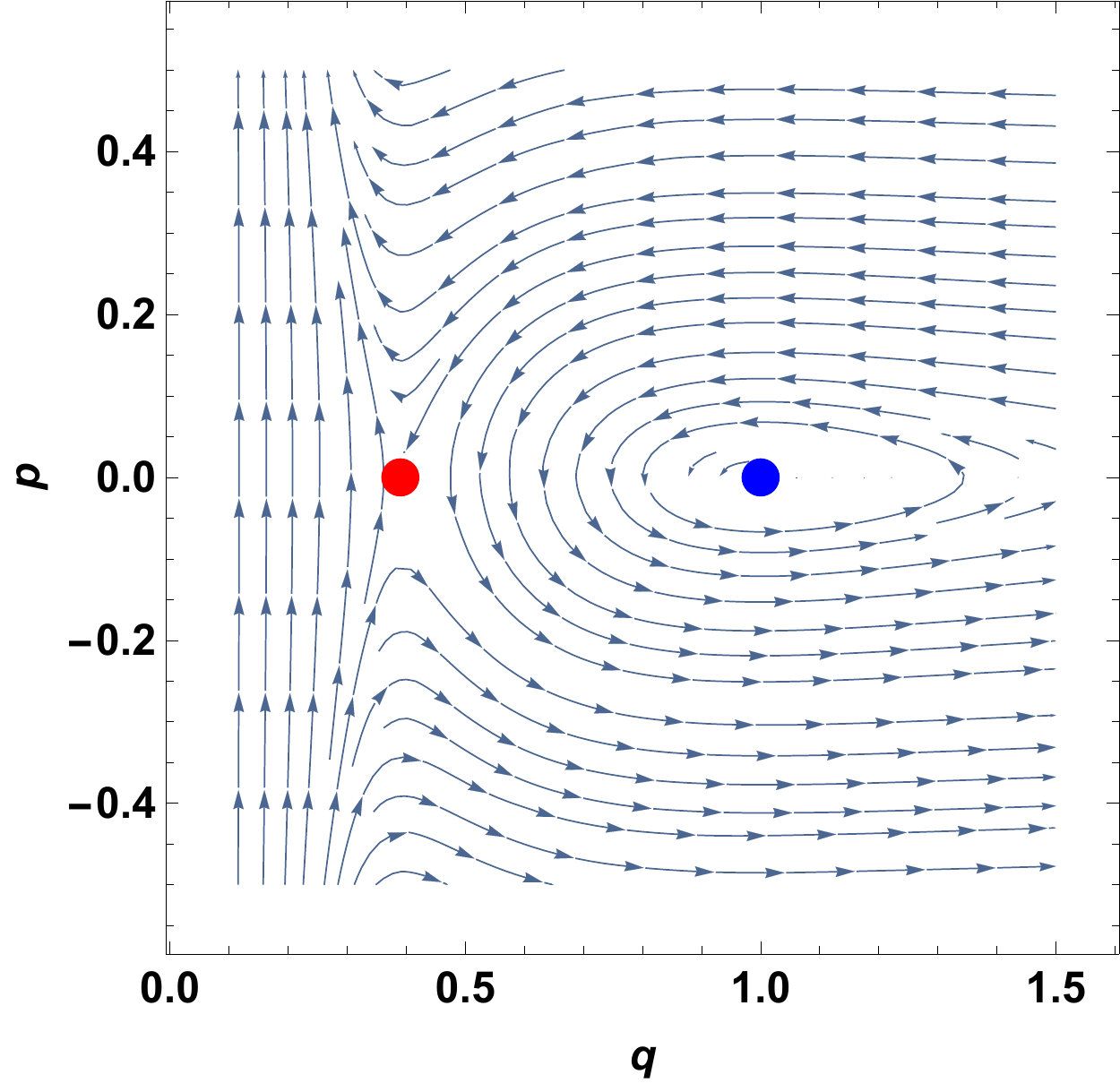}
\caption{Phase portrait in the canonical variables $(\qq,\pp)$ for $
  \alpha =3$ and a non local interaction kernel of range
  $\xi=.3 \Sigma^*$. Besides the elliptic fixed point at $q=1$ (blue dot) already present in the limit
  $\xi\to 0$, there appears a new (hyperbolic) fixed point at $q =
  O(\xi)$ (red dot) which
  governs the dynamics for optimization time long enough ($\mu \sigma^2 =1$).
  \sout{attract the dynamics for $\xi$ small enough}}
  \label{fig:portrait-smooth}
\end{figure}

\subsection{Non-Gaussian initial densities of agents}

In this last part of the section, we shall consider the situation
where we relax the assumption that the initial density of agents
$m_0(\bx)$ can be correctly described by a Gaussian. For sake of
clarity we limit this discussion to the one-dimensional case and local
linear interactions $\tilde V[m](x) = + g m(x)$.  We first consider
the situation where the initial density $m_0(\bx)$ can be described as
the juxtaposition of two well separated Gaussian-like bumps, and then
discuss some aspects of the general case.  Furthermore, we restrict
the discussion to the initial times (formation of the soliton),
assuming that final boundary conditions are compatible with a Gaussian
distribution.

To clarify the question of structured, but non Gaussian, initial
distributions of agents, it is useful to consider the example of an
initial condition which can be split into two well separated parts
$m_0(x) = m_0^a(x) + m_0^b(x)$, both separately well approximated by a
Gaussian and characterized by their relative masses, mean positions
and standard deviations, hereafter denoted by $\rho^k = \int m_0^k(x)
dx$, $X^k_0$ and $\Sigma^k_0$, with $k=a,b$, respectively.

We consider a variational ansatz which is a straightforward
generalization of
Eqs.~\eqref{eq:AnsatzGamma-1d}-\eqref{eq:AnsatzPhi-1d}, namely
\begin{align*}
\Phi(x,t) &=\Phi^a(x,t)+\Phi^b(x,t) \; ,\\
\G(x,t) &= \G^a(x,t) + \G^b(x,t) \; ,
\end{align*}
 where, for $k=a,b$,
\begin{align*}
\Phi^k(x,t) & = \sqrt{\rho^{k}}
 \exp \left[ \frac{- \gamma_t + P^{k}_t \cdot    x}{\mu\sigma^2}  \right] 
 \frac{1}{\left(2\pi(\Sigma^k_t)^2\right)^{1/4}} 
\exp \left[-\frac{(x- X^{k}_t)^2}{ (2\Sigma^{k}_t)^2} 
(1 - \frac{\Lambda^{k}_t}{\mu \sigma^2} )\right]
\\
\G^{k}(x,t) &= \sqrt{\rho^{k}}
 \exp \left[ \frac{+ \gamma_t - P^{k}_t \cdot    x}{\mu\sigma^2}  \right] 
 \frac{1}{\left(2\pi(\Sigma^k_t)^2\right)^{1/4}}
\exp \left[-\frac{(x- X^{k}_t)^2}{ (2\Sigma^{k}_t)^2} 
(1 + \frac{\Lambda^{k}_t}{\mu \sigma^2} )\right]
\end{align*}
and follow the same approach as in section~\ref{sec:variational-1}. We
make the two following assumptions: i) both parts $m_t^a(x)$ and
$m_t^b(x)$ remain well separated for the time necessary to get an
equilibrium shape, $|X^a_t - X^b_t| \gg \ \Sigma^{a}_t+
\Sigma^{b}_t$, and ii) the extensions $\Sigma^{a}_t$, $\Sigma^{b}_t$
of both parts are small on the scale at which $U_0(x)$ varies
significantly.

Under these assumptions, the time evolution greatly simplifies as each
sub-part behaves independently form the other and follow
Eqs.~\eqref{eq:XdotP}-\eqref{eq:PdotP} for its center of mass, and
Eqs.~\eqref{eq:Sigma-dot-1d}-\eqref{eq:Lambda-dot-1d} for the
distribution parameters, with an effective coupling constant $g^k=
\rho^k g$ in the interaction potential.

Thus, using the results of the previous section, we get that each
sub-part $k$ forms a soliton with a rescaled extension $\Sigma^k_*
=\frac{1}{\rho^k} \Sigma_*$ (with $\Sigma_*$ as in \eqref{eq:fp-11b}),
which implies that the smaller part gets the larger extension.  This
first evolution takes place on time scales $\tau^k_* = \sqrt{\rho^k}
\tau_*$ (see Eq.~\eqref{eq:tau*-1d} and the discussion at the end of
section~\ref{sec:TimeEvolSigmaLambda}).

Let us now analyze the separate motion of the centers of mass for each
part, $(X^{k}_t,P^{k}_t)$. In order to keep with the simplest picture,
we completely neglect the effect of $U_0(x)$ (relaxing this
  assumption does not introduce major conceptual changes) and suppose
that in the long time limit, the density of agents should form a
single soliton of mass one and at rest. In such a case, the solitons
evolve as independent classical particles with unknown constant
velocities $v^{k} =P_0^{k}/\mu$, which have to be determined.

Conservation of total energy and the final condition chosen implies
that the system of the two initially separated solitons have the same
energy as a single soliton at rest, that is
\begin{equation}
\sum_{k=a,b} \rho^k \left\{ \frac{1}{2} \mu [v^k]^2 + [\rho^k]^2
  \tilde E_{\rm tot}^*\right\} =  \tilde E_{\rm tot}^* 
\end{equation}
where we have used that the energy of a soliton in the center of mass
$\tilde E_{\rm tot}^*$ (see Eq.~\eqref{eq:Etotstar-1d}) scales as the
square of the coupling constant $g$. Then, in the absence of an
external potential $U_0$, total momentum is conserved (cf
Eq.~\eqref{eq:Pdot}),
\begin{equation}
\sum_{k=a,b} \mu\, \rho^k v^k= 0 \; .
\end{equation}
The velocities of the solitons before collision are thus
\begin{equation}
|v^k|= (1-\rho^k)\sqrt{\frac{6  \tilde E_{\rm tot}^*}{\mu}} = \sqrt{\frac{3}{4\pi}} (1-\rho_k) v_g \; , \label{eq:vk} 
\end{equation}
for $k=a,b$ and with $v_g$ defined by Eq.~\eqref{eq:vg}, the velocity scale associated
with the interactions.

If the pair of solitons have an extension initially larger than their
invariant value $\Sigma_*^k$, they contract with the initial velocity
given by Eq.~\eqref{eq:dSigma2-1d},
\begin{equation} \label{eq:sigma-dot-ab}
  \dot \Sigma^{a,b}_t \simeq - \frac{\Sigma_*^k}{\tau_*} =  
  - \frac{1}{\sqrt{4\pi}} \rho^{k}  v_g  \; .
\end{equation}
We find that light solitons have a contraction velocity slower than
their center of mass, resulting in a positive velocity of the front of
matter toward the other soliton before they reach their equilibrium
shape.  However, equilibration time is smaller for lighter solitons,
$\tau_c^k= \rho^k \frac{\Sigma_0}{\Sigma_*}\tau_*$, so that the front
of matter, $X^k_t+\Sigma^k_t$ moves by a finite fraction of $\Sigma_0$
in the time necessary for $\Sigma^k_t$ to reach the value $\Sigma^k_*$
(the maximal value over $\rho^k$ is found to be $\frac{3}{4
      (1+\sqrt{3})} \Sigma_0 \simeq .27 \Sigma_0$). Thus the picture
we gave here is consistent with the hypothesis of an initial
separation, $\Sigma^{a}_0 + \Sigma^{b}_0\ll |X_0^a - X_0^b|$.

For arbitrary initial conditions, the exact scenario may become
significantly more complex, and a precise description which would be
universally valid is obviously beyond the scope of the present work.
We limit ourselves to what can be anticipated on a general basis.

The case of two solitons studied above can easily be generalized to a
larger number: if the initial density of agents $m_0(x)$ can be
separated in a few non-overlapping sub-part of mass $\rho^k$ and size
larger than $({\Sigma_*}/{\rho^k})$, each of these sub-parts
contracts and forms a local soliton of extension inversely
proportional to its mass which moves until it merges with a
neighboring soliton.  Furthermore, if the size $\rho^k$ of these
sub-parts is uniformly bounded from below, then the formation of local
solitons is characterized by the velocity scale $v_g$ and occurs on a
short time scale in the limit of strong interactions.  In this
setting, lighter solitons take more time to form, and move faster than
the heavier ones.  When more than two solitons are present, various
scenarios are possible which differ by the order in which they merge
together, implying different choices of initial conditions.

  More general initial densities of agents with inhomogeneities but no
  clearly separated sub-parts would have to be studied in a case by
  case basis.  However, the fact that the extension of local solitons
  is inversely proportional to their mass lead us to expect the
  formation of solitons for strong enough interaction potential, even
  if the determination of their distribution would remain a-priori a
  difficult problem.

\section{Perturbative approach to the weakly interacting regime}
\label{sec:perturbative}
\newcommand{\gauss}{\mathsf{G}}
\newcommand{\losc}{\ell_{\rm osc}}
\newcommand{\ho}{{H_0}}

In contrast with the two previous sections, we now turn to the case
when the interactions between agents are small with respect to the
external potential so that they can be  described as a perturbation of a
non-interacting model.  

The general strategy here is relatively clear.  If the ``interaction
potential'' $V[m](x)$ is small, one should first solve the
(non-interacting)  Schr\"odinger equation (as in
section~\ref{sec:free-case}); we then plug in the interactions,
assumed to be small, and insert the potential term $V[m](x)$ as a
perturbation, using the standard tools of quantum mechanics.   We
shall see however that the forward/backward structure of the Mean
Field Game equations introduces some subtleties in this relatively
straightforward scenario.

Here, we limit ourselves to the description of interactions up to
first order corrections.  Furthermore, we shall consider here the long
horizon limit $T \to \infty$, and concentrate mainly on the
convergence to the ergodic state.  We thus discard the effects of the
final boundary conditions, which show up only in the late stage of the
process.

As an application, we will consider  a one-dimensional model 
with a quadratic (inverted) external potential
\begin{equation} \label{eq:U0}
U_0(x) = - \frac{k}{2}  x^2 
\end{equation}
 and a weak short-ranged interaction potential 
\begin{equation} \label{eq:WeakInter}
 V[m](x) = g\, m(x)
\end{equation}
where  $g$ is a small positive coupling constant.

\subsection{Non-interacting model}

We start with a brief discussion of the non-interacting limit $\tilde
V[m](x) \equiv U_0(x)$, mainly to fix some notations and to recall some
properties we shall make use of in this section.

From the results of section~\ref{sec:free-case}, we know that the time
evolution of both functions $\Phi$ and $\G$ can be derived from the
eigenfunctions $\psi_n(x)$ and eigenvalues $\lambda_n$ of the
Hamiltonian
\begin{equation}
 H_0 = -  \frac{1}{2\mu} \hat \Pi^2 - U_0(x)  \; .\label{eq:H0-unperturbed}
 \end{equation}
The time evolution of the two functions $\Phi(x,t)$ and $\G(x,t)$ can be
expressed in terms of these eigenfunctions through the construction of a
propagator
\begin{equation} \label{eq:prop-def}
G_{\ho }(x,x',t) \equiv \sum_{n \geq 0} e^{- \lambda_n \, t/ \mu \sigma^2}
\psi_n(x) \psi_n(x') \; ,
\end{equation}
where the subscript $H_0$ stands for the ``free'' Hamiltonian \eqref{eq:H0-unperturbed}.
We have 
\begin{align}
\Phi(x,t) & =  \int dx\, \Phi(x',t')\, G_{\ho }(x',x,t'-t) \qquad t \le t' \\
\G(x,t)    & = \int dx' \,G_{\ho }(x,x',t-t')\, \G(x,t')   \qquad t \ge t' \label{eq:PropPhi}
\end{align}

We now consider the influence of both the initial density of agents
$m_0(x) =  \Phi(x,0)\G(x,0)$ and the terminal condition $\Phi(x,T) = K
\exp \left(-c_T(x)/\mu \sigma^2\right)$. In the long horizon limit
$T \gg \terg$, where we define the ergodic time as  
\begin{equation} \label{eq:terg}
\terg \equiv \frac{\mu \sigma^2}{\lambda_1-\lambda_0} \; ,
\end{equation}
the system gets close to the ergodic state at all intermediate times
$t$ such that  $t \gg \terg$ and $(T-t) \gg \terg$. The terminal
condition becomes thus irrelevant, except possibly in the late stages that
we do not consider here.  The ergodic state in the absence of interactions is
\begin{align} 
\Phi_e^{\ho }(x,t) & \equiv \; C\; e^{+ \lambda_0 t / \mu\sigma^2} \psi_0(x)  \label{eq:phie} \\
\G_e^{\ho }(x,t)   & \equiv  C^{-1} e^{- \lambda_0 t / \mu\sigma^2} \psi_0(x)  \label{eq:gammae} 
\end{align}
(with $C$ some arbitrary constant that we  fix to one here), so that the resulting density profile is time independent, $m(x,t)
= m_e^\ho(x) = \psi^2_0(x)$.  The backward time evolution of $\Phi(x,t)$
coming from the ergodic state $\Phi_e^{\ho }$ at some fixed final time is trivial, and in particular $\Phi(x,0) = \psi_0(x) $.  The initial
condition for $\G$ thus reads
\begin{equation}
\G(x,0) = \frac{m_0(x)}{\psi_0(x)}
\end{equation}
and  for all further times $t$ 
\begin{equation}
\G(x,t) = \int dx' G_{\ho }(x,x',t)
\frac{m_0(x')}{\psi_0(x')} . \label{eq:Gamma-prop} 
\end{equation}
Since $m(x,t) = \Phi(x,t)\G(x,t)$ (cf \eqref{eq:Gamma}), the time
  evolution for the density  can be written as
\begin{equation} \label{eq:density-prop}
m(x,t) = \int dx' F_{\ho }(x,x',t) m_0(x') 
\end{equation}
where we have introduced the density time-propagator
\begin{equation} \label{eq:Fdef} 
 F_{\ho }(x,x',t)   \equiv \psi_0(x) G_{\ho }(x,x',t)\frac{e^{+ \lambda_0 t /\mu\sigma^2}}{\psi_0(x')}  
  \; . 
\end{equation}
As stressed before, these expressions for the propagation of the
density of agents are valid in the long optimization time limit
$T \gg  \terg$, and their simplicity can be
eventually traced back to the fact that $\Phi(x,t)$ remains in its
ergodic state $\Phi_e$ as long as $ (T-t) \gg \terg$, and in
particular near $t=0$. 

We end this subsection with a few comments.  First, we stress that for
times large enough, the propagator \eqref{eq:prop-def} factorizes
\begin{equation} \label{eq:G-large-asymp}
G_{\ho }(x,x',t ) \simeq  e^{- \lambda_0 t /\mu\sigma^2} \psi_0(x) \psi_0(x') \qquad \hbox{for all } t\,\gg\,\terg,
\end{equation}
and one recovers as expected the ergodic state $\G(x,t) = \G_e(x,t)$
from \eqref{eq:Gamma-prop}, and $m(x,t) = m_e^H(x)$ from
\eqref{eq:density-prop}.  This implies in particular that for any
normalized initial density $m_0(x)$
\begin{equation} \label{eq:F-large-asymp}
 \int dx' F_{\ho }(x,x',t) m_0(x') \simeq  m_e(x) \qquad  \hbox{for all } t\,\gg\,\terg .
\end{equation}
Finally,  one can check easily that $m_e$ is  a fixed point of  the
propagation equation~\eqref{eq:density-prop},
\begin{equation}
m_e^{\ho }(x) = \int dx' F_{\ho }(x,x',t) m_e^{\ho }(x')  \; ,
\end{equation}
as again expected.  We shall make use of these properties
below.

\subsection{First order perturbations :  the ergodic state}
\label{sec:ergo-pert}
\newcommand{\tc}{{t^*_c}}

We want now to compute the first order corrections to the previous
non-interacting model when a weak interaction term $ V(x,t) = g\,
m(x,t)$ is added to the potential.

Let us denote by $\psi_e$ the solution of the nonlinear ergodic problem,
\begin{equation} \label{eq:erg-pro-pert}
  \lambda_e \psi_e(x)  = - \frac{\mu\sigma^4}{2} \Delta\psi_e(x) - U_0(x)
\psi_e(x) -g \left(\psi_e(x)\right)^3  \; .
\end{equation}
For $g$ small enough, both $\psi_e$ and $\lambda_e$ can be computed
using perturbation theory around the lowest energy state of $H_0$ (Eq.~\eqref{eq:H0-unperturbed}). To first order in $g$, one gets easily
\begin{align} \label{eq:psierg-approx}
\psi_e(x) &= \psi_0(x) + g \sum_{n>0} \frac{V_{0,n}}{(\lambda_n -
  \lambda_0)}  \psi_n(x) \; +o(g) \\
\lambda_e &= \lambda_0 - g\, V_{0,0} \; +o(g) \label{eq:lambdaerg-approx}
\end{align}
where $\psi_n$ is the unperturbed eigenfunction of $H_0$
and for all $n,m \ge 0$,
\begin{equation}\label{eq:Vnm}
 V_{n,m} \equiv \int dx \, \psi^2_0(x) \psi_n(x) \psi_m(x)  \; .
 \end{equation}
The density in the ergodic state then reads, up to first order,
\[
m_e(x) = \psi_0^2(x) + 2 g\, \psi_0(x) \sum_{n>0} \frac{V_{0n}}{
  (\lambda_n - \lambda_0)} \, \psi_n(x) \; +o(g) \; .
\]

\subsection{First order perturbations : dynamics}

We now construct the dynamic evolution toward the ergodic state, given
an initial density profile $m_0$. 

The basic tool we shall use in this subsection is essentially the time
dependent perturbation theory of quantum mechanics.  However the
forward/backward structure of the mean field game equations introduces
some extra complication since, as we shall see, it requires to have
perturbative results which remain valid for very long times (of the
order of the optimization time $T$).

For this reason, it turns out to be necessary to develop the
perturbation theory not around the unperturbed Hamiltonian $H_0$
(Eq.~\eqref{eq:H0-unperturbed}) but around the Hamiltonian
associated with the true ergodic state
\begin{equation}\label{eq:Ham-erg}
H_e(x) = -\frac{\Pi^2}{2\mu} -U_0(x) -g
\left(\psi_e(x)\right)^2 \; ,
\end{equation}
where for now $\psi_e(x)$ denotes the exact solution of the ergodic
problem Eq.~\eqref{eq:erg-pro-pert}.  We then write the full time
dependent Hamiltonian as $H_e$ plus a perturbation:
\begin{equation}
H(x,t)= H_e(x) - g \left( m(x,t) - m_e(x)\right)
\end{equation}
where $m_e(x)$ is the (assumed known) ergodic density of state and
$m(x,t)$ is the yet unknown time dependent density of agents.

Let $T_e$ be a time at which the system is in the ergodic state
(possibly up to an exponentially small error in $T$ that we fully
neglect here). $\Phi(x,t)$ is solution of
\begin{equation}
\mu\sigma^2\,\partial_t \Phi(x,t) = H(x,t) \Phi(x,t)
\end{equation}
with the terminal condition
\begin{equation}
\Phi(x,T_e) = e^{ \lambda_e T_e /\mu\sigma^2} \psi_e (x) \; .
\end{equation} 
Following standard time-dependent quantum perturbation theory
\cite{book:Sakurai}, $\Phi(x,t)$ reads, up to first order in the
perturbation and for all $t$ in $[0, T_e]$,
\begin{align*}
\Phi(x,t) &= e^{ \lambda_e t /\mu\sigma^2}  \psi_e(x) \\
&\quad+    \frac{g}{\mu\sigma^2}\int_t^{T_e} ds
    \int dy \,  \psi_e(y) \, e^{(\lambda_e s /\mu\sigma^2)  }  [m(y,s) - m_e(y)]  G_e(y,x,s-t) 
\end{align*}
where we have denoted by $G_e(y,x,t)$ the propagator
\eqref{eq:prop-def} associated with $H_e$.  

In analogy with Eq.~\eqref{eq:Fdef}, we introduce the density
time-propagator associated with $H_e$,
\begin{equation} \label{eq:properg}
F_e(y,x,t) = {\psi_e(y)}\, G_e(y,x,t) e^{+ \lambda_e t/\mu\sigma^2  }
\, \frac{1}{\psi_e(x)} \; ,
\end{equation}
 and we  write the evolution of $\Phi$ as
\begin{equation}
\Phi(x,t) = e^{\lambda_e t/\mu\sigma^2 } \left[
1 + \frac{g}{\mu\sigma^2}\int_t^{T_e} ds \int dy\,
[m(y,s) - m_e(y)]   F_e(y,x,s-t) \right] \psi_e(x)  \; .
\end{equation}
Note that in this last expression, the reasons for the choice of a
perturbation theory around $H_e$ rather than around $H_0$ can be made
clear: first, the exponential factor may differ greatly from the same
expression with $\lambda_0$ instead of $\lambda_e$, since $t$ can get
large independently of $g$; second, with the present choice, the time
integral in the right hand side is  well defined, even in the limit
$T_e \to \infty$ (assuming the convergence of the expansion). 

In particular, the value at $t=0$ reads
 \begin{equation}
\Phi(x,0)  = \left[
1 + \frac{g}{\mu\sigma^2}\int_0^{T_e} ds \int dy\,
[m(y,s) - m_e(y)]   F_e(y,x, s) \right] \psi_e(x) 
\end{equation}
Given an initial distribution of agents $m_0$, the initial value of
$\Gamma(x,0)$ can be now computed up to first order as 
 \begin{align}
   \Gamma(x,0) &=  \left[ 1 -
     \frac{g}{\mu\sigma^2}\int_0^{T_e} ds \int dy\, [m(y,s) - m_e(y)]
     F_e(y,x, s) \right] \frac{m_0(x)}{\psi_e(x)}
      \end{align}

Now,  since $\Gamma(x,t) $ is solution of 
\begin{equation}
- \mu\sigma^2\,\partial_t \Gamma(x,t) = H(x,t) \Gamma(x,t)
\end{equation}
it can be written up to first order in $g$ as
\begin{align*}
\Gamma(x,t) &= \int dx'  G_e(x,x',t) \Gamma(x',0) \\
&  + \frac{g}{\mu\sigma^2}\int_0^t ds \int dx' dy \,G_e(x,y,t-s)
 [m(y,s) - m_e(y)]  G_e(y,x',s) \Gamma(x',0)    \;.
\end{align*}

Collecting the previous results, one can write an expression for
the density of agents at first order in perturbation theory. 
We get 
\begin{align}
m(x,t) &= \int dx' F_e(x,x',t) \,m_0(x') \nonumber\\
& + \frac{g}{\mu\sigma^2} \int_0^{t} ds \int  dy \,	dx'	 
 F_e(x,y,t - s) \left[m(y,s)-  m_e(y) \right]
F_e(y,x',s)  m_0(x')  \nonumber\\
&+\frac{g}{\mu\sigma^2} 
 \Bigl[\int_t^{T_e}  ds \int dy \left[m(y,s)-  m_e(y)\right]F_e(y,x,s-t) \Bigr]  \Bigl[ \int dx' F_e(x,x',t) m_0(x') \Bigr]
  \nonumber \\
  & - \frac{g}{\mu\sigma^2} \int_0^{T_e} ds \int  dy dx' F_e(x,x',t)
  \left[m(y,s)-  m_e(y) \right]   F_e(y,x',\,s\,)    
 m_0(x')   \label{eq:mpert-v1} 
\end{align} 

Note that there are thus three terms at the first order of
perturbations with a different origin: though the two last terms terms
are rather classical and correspond to the first order perturbation of
each of the two factors $\Gamma(x,t)$ and $\Phi(x,t)$, respectively,
the third one is specific to the forward/backward structure and
corresponds to a modification of the initial data $\Gamma(\cdot,0)$.

As a coherence check of the above expression, we can consider the long
time limit $t \to \infty$, of Equation \eqref{eq:mpert-v1} and
verify that, at first order in $g$, it is indeed coherent with the
expected result that $m(x,t) \simeq
m_e(x)$ whenever $t \! \gg \! \terg$.  Using the fact that at the
lowest order, $|m(x,t)-m_e(x)|\le C e^{-t/\terg}$, (which we write as
$m(x,t)\simeq m_e(x)$), we get from Eq.~\eqref{eq:G-large-asymp}, at
first order in $g$
\begin{align}
m(x,t) &\simeq \int dx' F_e(x,x',t )  m_0(x')\nonumber\\ 
& + \frac{g}{\mu\sigma^2} \int_0^{t} ds \int  dy \,	dx'	 
 F_e(x,y,t - s) \left[m(y,s)-  m_e(y) \right]
F_e(y,x',s)  m_0(x')  \nonumber\\
  & - \frac{g}{\mu\sigma^2} \int_0^{t} ds \int  dy dx' F_e(x,x',t)
  \left[m(y,s)-  m_e(y) \right]   F_e(y,x',\,s\,)    
 m_0(x')  \nonumber\\
 &\simeq \int dx' F_e(x,x',t )  m_0(x')\nonumber
\\ & + \frac{g}{\mu\sigma^2} \int_0^{t} ds \int  dy \,	dx'	 
 \left[F_e(x,y,t - s) -F_e(x,x',t) \right]\nonumber \\
 &\qquad \quad\qquad\times\left[m(y,s)-  m_e(y) \right]
F_e(y,x',s)  m_0(x')  \nonumber\\
&\simeq m_e(x)\ \label{eq:m-large-t}
  \end{align}
  where we have used also the fact that $F_e(x,y,t) \simeq m_e(x) $ to
  get the last line, valid up to first order in $g$.  For short times,
  $t\le \mu\sigma^2/ |V_{1,1}-V_{0,0}|$ (assuming that
  $V_{1,1}\not=V_{0,0}$), the dynamics towards the ergodic state does
  not differ too much from the unperturbed one and the densities and
  propagators in the right hand side of Equation \eqref{eq:mpert-v1}
  can be replaced by their first order approximations in the first
  line, and their expressions at $g=0$ in the next three lines. We
thus get an explicit form the first order solution to the Mean
Field Game equations:
\begin{align}
    m(x,t) &=m_e^{\ho }(x)+ \int dx' F_{\ho }(x,x',t) \,(m_0(x') -m_e^{\ho }) \nonumber\\
    &+ \int dx'(F_e(x,x',t) - F_{\ho }(x,x',t)) \,m_0(x') \nonumber\\
& + \frac{g}{\mu\sigma^2} \int_0^{t} ds \int  dy \,	dx'	 
\left[ F_{\ho }(x,y,t - s) - F_{\ho }(x,x',t)\right] \nonumber\\
&\qquad\qquad\qquad\times \left[m^{\ho }(y,s)-  m_e^{\ho }(y) \right]
F_{\ho }(y,x',s)  m_0(x')  \nonumber\\
&+\frac{g}{\mu\sigma^2} 
  \int_t^{T_e}  ds \int dy\,dx' \left[m^{\ho }(y,s)-  m_e^{\ho }(y)\right] \nonumber\\
 & \qquad\qquad\qquad\times \left[F_{\ho }(y,x,s-t)- F_{\ho }(y,x',\,s\,) \right]   F_{\ho }(x,x',t) m_0(x') 
\label{eq:mpert-v2}
  \end{align}
where the term on the second line accounts for the first order correction to the propagator.

For large times, ($t \gg \terg$), this expression converges
exponentially to $m_e(x)$ for the same reasons as in
Eq.~\eqref{eq:m-large-t}.

For short times,
($t \ll \terg$), it leads to $m(x,t)  = m_0 (x) + \partial_t m(x,0)\,
t + O\left((t/\terg)^2 \right)$, where, within first  order
approximation,
\begin{equation} \label{eq:dm-short-t}
 \partial_t m(x,0) \simeq \partial_t m^{(0)}_0 (x) +
 \frac{g}{\mu\sigma^2}  \partial_t m^{(1)}_0 (x) 
\end{equation}
with 
\begin{equation} \label{eq:mt00}
\partial_t m^{(0)}_0 (x) = \int dx'\partial_t F_{\ho}(x,x',0) m_0(x') 
\end{equation}
the ``free'' contribution.  We can furthermore write the first order
correction as the sum
\begin{equation}\label{eq:intdec}
\partial_t m^{(1)}_0 (x) = \partial_t m^{(e)}_0 (x)+\partial_t m^{(a)}_0 (x)+\partial_t m^{(r)}_0 (x)
\end{equation}
with
\begin{align}
 \partial_t m^{(e)}_0 (x) =& \left( \frac{g}{\mu\sigma^2}\right)^{-1}
 \int dx' [\partial_t F_e(x,x',0)- \partial_tF_{\ho}(x,x',0)] m_0(x')  \\ 
 \partial_t m^{(a)}_0 (x)=& - m_0(x)
\int_0^{\infty}  ds \int dy \delta m^{\ho }(y,s)  
\partial_t  F_{\ho }(y,x,s)  \\ 
  &  + \int_0^{\infty} ds \int dy\,dx' 
  \delta m^{\ho }(y,s) 
 F_{\ho }(y,x,s)  \partial_t
  F_{\ho }(x,x',0) \,m_0(x')  \\
\partial_t m^{(r)}_0 (x)=& -  \int_0^{\infty} ds \int dy\,dx'   \delta
 m^{\ho}(y,s)  F_{\ho }(y,x',s)  \partial_t  F_{\ho }(x,x',0)
 \, m_0(x')  \; .
\end{align}
In the expressions above we have introduced the notation $\delta
m^{\ho }(y,s) \equiv [m^{\ho }(y,s) -m^{\ho }_e(y) ]$, and the spatial
integrals for $\partial_t m^{(0)}_0$, $\partial_t m^{(e)}_0$,
$\partial_t m^{(a)}_0$ and $\partial_t m^{(r)}_0$, can be further
simplified using that
\[
\partial_t F_{\ho}(x,y,0) = \frac{\sigma^2}{2} \left(\psi_0(x)
  \delta"(x-y) - \psi"_0(x)  \delta(x-y)  \right) \frac{1}{\psi_0(y)}
\; ,
\]
(and the equivalent expression for $\partial_t F_{e}(x,y,0)$).

The term $\partial_t m^{(e)}_0$ can be understood as originating from
the modification of the {\em ergodic state} (cf
section~\ref{sec:ergo-pert}), $\partial_t m^{(a)}_0$ is related to the
influence of interactions on {\em anticipations} and $\partial_t
m^{(r)}_0$ is due to the more (in time) classical, retarded effect of
interactions.

We now apply these results to the example of the harmonic oscillator
potential Eq.~\eqref{eq:U0}, for which we compute the first order,
$O({g}/{\mu\sigma^2})$, corrections to $\partial_t m(x,0)$.

\subsection{Weakly interacting agents in an harmonic potential}

We consider now in more details the harmonic case 
\[ U_0(x) = - \frac{k}{2} x^2 \; . \]
Our goal here is to obtain explicit expressions for the various
quantities involved in Eq.~\eqref{eq:mpert-v2}, and more specifically
the density propagator $F_{\ho }$ and the time-dependent density
of agents $ m^{\ho }(x,t)$.

When the potential is harmonic  the eigenfunctions  of the unperturbed
Hamiltonian \eqref{eq:H0-unperturbed} can be written as 
\begin{equation} \label{eq:varphi-n}
 \psi_n(x) =  \frac{1}{\sqrt{2^n n !}} \frac{1}{\pi^{1/4} \sqrt{\ell_0}}
\exp\left(-  \frac{x^2}{2 \ell_0^2} \right) 
H_n\left( \frac{x}{\ell_0}\right) \; ,\quad n\ge 0
\end{equation}
where $\ell_{\rm o} = \sigma \left(\frac{\mu}{k}\right)^{1/4}$ and
$H_n(u)$ is the $n^{\rm th}$ Hermite polynomial; the associated
eigenvalues are $ \lambda_n = \mu \sigma^2 \omega (n + 1/2)$ with
$\omega=\sqrt{k/\mu}$.  In particular the ground state of $H_0$ reads
\begin{equation} \label{eq:hc-gs}
\psi_0(x) = \frac{1}{\pi^{1/4} \ell_{\rm o}^{1/2}}
\exp\left(- \frac{1 }{2} \frac{x^2}{\ell_{\rm o}^2} \right) 
\; .
\end{equation}
By Mehler formula (see reference \cite{WATSON} or
Appendix~\ref{app:perturbative} for a derivation) the propagator
Eq.~\eqref{eq:prop-def}  reads explicitly
\begin{align} \label{eq:prop-explicit}
G_{\ho }(x,x',t) & = 
\frac{1}{\sqrt{2\pi \ell_0^2 \sinh(\omega t)}}
\exp \left[ - 
\frac{ (x^2+{x'}^2) \cosh(\omega t)  - 2xx' }{2 \ell_0^2 \sinh(\omega
  t)}\right] \\
& \simeq \frac{e^{- \omega t/2}}{\sqrt{\pi } l_0} \exp \left[ - 
\frac{ (x^2+{x'}^2) )  }{2 \ell_0^2 }\right] \qquad \mbox{when $(\omega t \gg 1)$}
 \nonumber
\end{align}
This in turns implies that 
\begin{equation}
 F_{\ho }(x,x',t)   = \gauss_{\Sigma_F(t)}(x -x' e^{-\omega  t}) \; , \label{eq:Fgauss}  
\end{equation}
with
\begin{equation}
\Sigma_F(t) = \ell_0\sqrt{(1 - e^{-   2 \omega t})/2}  \; , \nonumber
\end{equation}
where  $\gauss_\Sigma$ is a centered Gaussian of width $\Sigma$ 
\begin{equation}
\gauss_\Sigma(x) =  \frac{1}{\sqrt{2\pi}\Sigma} \exp \left[-
  \frac{x^2}{2 \Sigma^2}\right] \; .
\end{equation}
Using Eq.~\eqref{eq:Fgauss}, the implementation of
Eq.~\eqref{eq:mpert-v2}, beyond the zero'th order term, now reduces to
quadratures.

In the particular case of a Gaussian initial
condition 
\begin{equation} \label{eq:gauss-m0}
m_0(x) \equiv \gauss_{\Sigma_0}(x-x_0)
\end{equation} 
the integration in Eq.~\eqref{eq:density-prop} can be performed and
the time dependent density profile is Gaussian at all times, with
\begin{equation}
m^{\ho }(x,t)   = \gauss_{\Sigma_m(t)}(x- x_0\, e^{-\omega t}) \label{eq:gauss-mt} 
\end{equation}
with
\begin{equation}
\Sigma_m(t)   = \ell_0\sqrt{\left(1 - (1-2\, \Sigma_0^2/\ell_0^2)\,
    e^{-   2 \omega t}\right)/2} \; .
\end{equation}

For time small enough, the density-propagator $F_{\ho }(x,x',t)$ is peaked
around $x' = x e^{\omega t}$, and the integral in
Eq.~\eqref{eq:density-prop} is dominated by a neighborhood of size
$e^{\omega t}\,\Sigma_F(t)$ around this value.  If the initial
profile $m_0(x')$ is slowly varying on this length-scale,  the corresponding term can be factorized out
  of  the integral in Eq.~\eqref{eq:density-prop} (which is akin to
performing a stationary phase approximation), leading to a simpler
expression for the density at short times
\[
m(x,t) \simeq e^{\omega t} m_0\left(x e^{\omega t} \right) \; .
\]
(Note that this does not require $m_0$ to be a Gaussian.)
 If the typical scale of variations $\Sigma_0$  of  the initial
 distribution of agents is significantly 
larger than the length $\ell_0$ characterizing the ground state
$\psi_0(x)$, this 
approximation will be valid up to time $t \sim \omega^{-1}
\log(\Sigma_0/\ell_0)$.

We turn now to the short time evolution of the density for a Gaussian
initial density as in \eqref{eq:gauss-m0}.  In
expression~\eqref{eq:dm-short-t} all integrations over space are
Gaussian and can be made explicitly, which leaves only the
integration over time to be performed numerically. The result up to
first order in the interaction strength is illustrated in
Fig.~\ref{fig:shorttime} for a particular set of parameters (see
caption for more details).

\begin{figure}[t]
\includegraphics[width=6.5cm,clip]{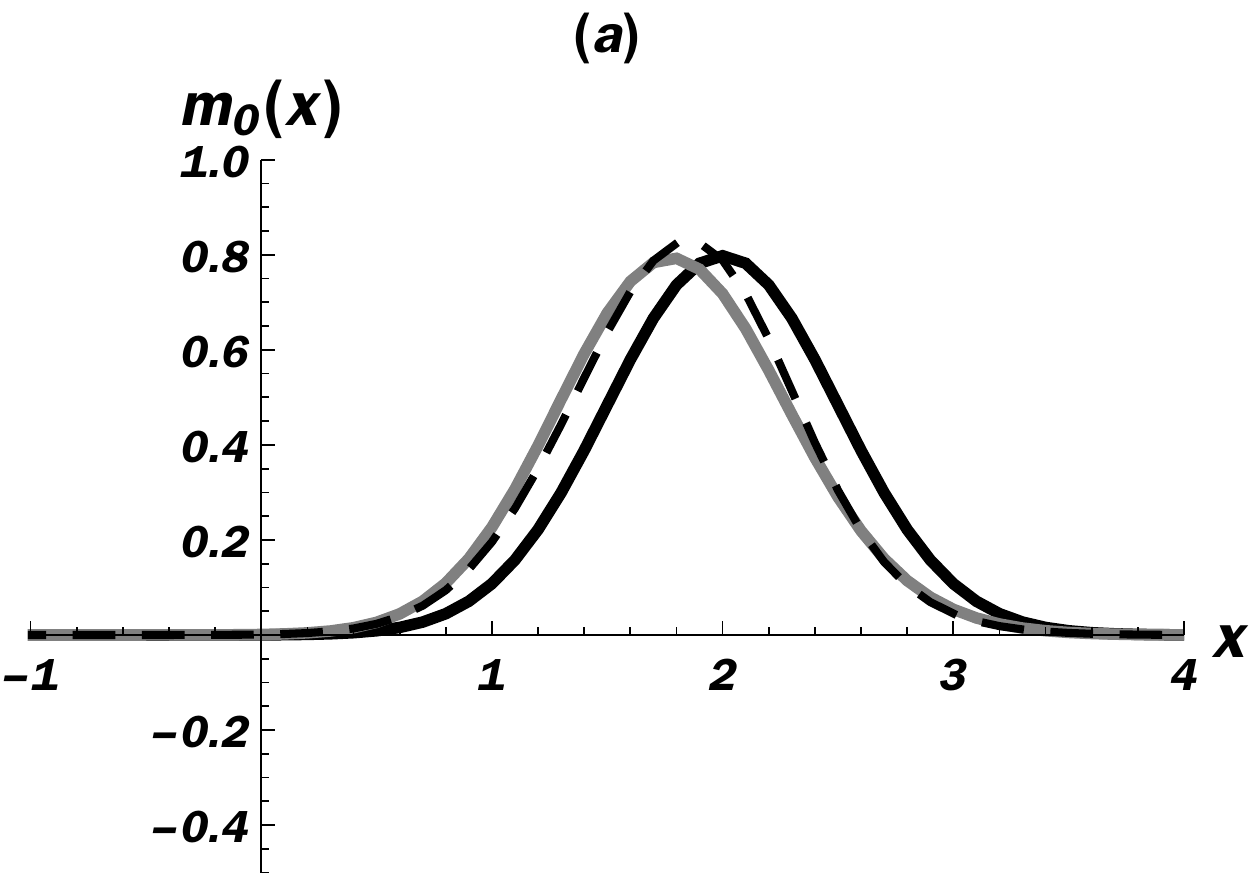}
\hspace{.5cm} 
\includegraphics[width=6.5cm,clip]{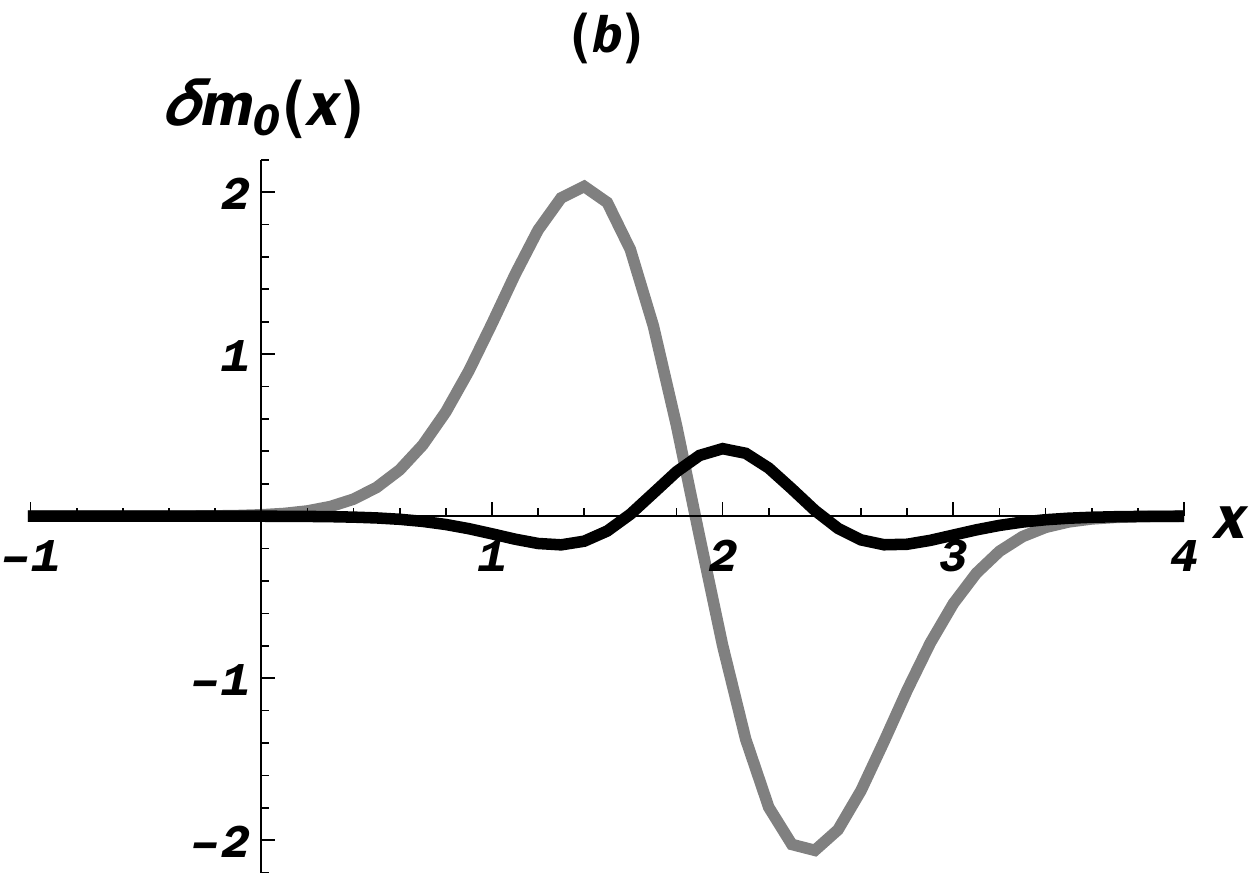}
\hspace{.5cm} 
\includegraphics[width=6.5cm,clip]{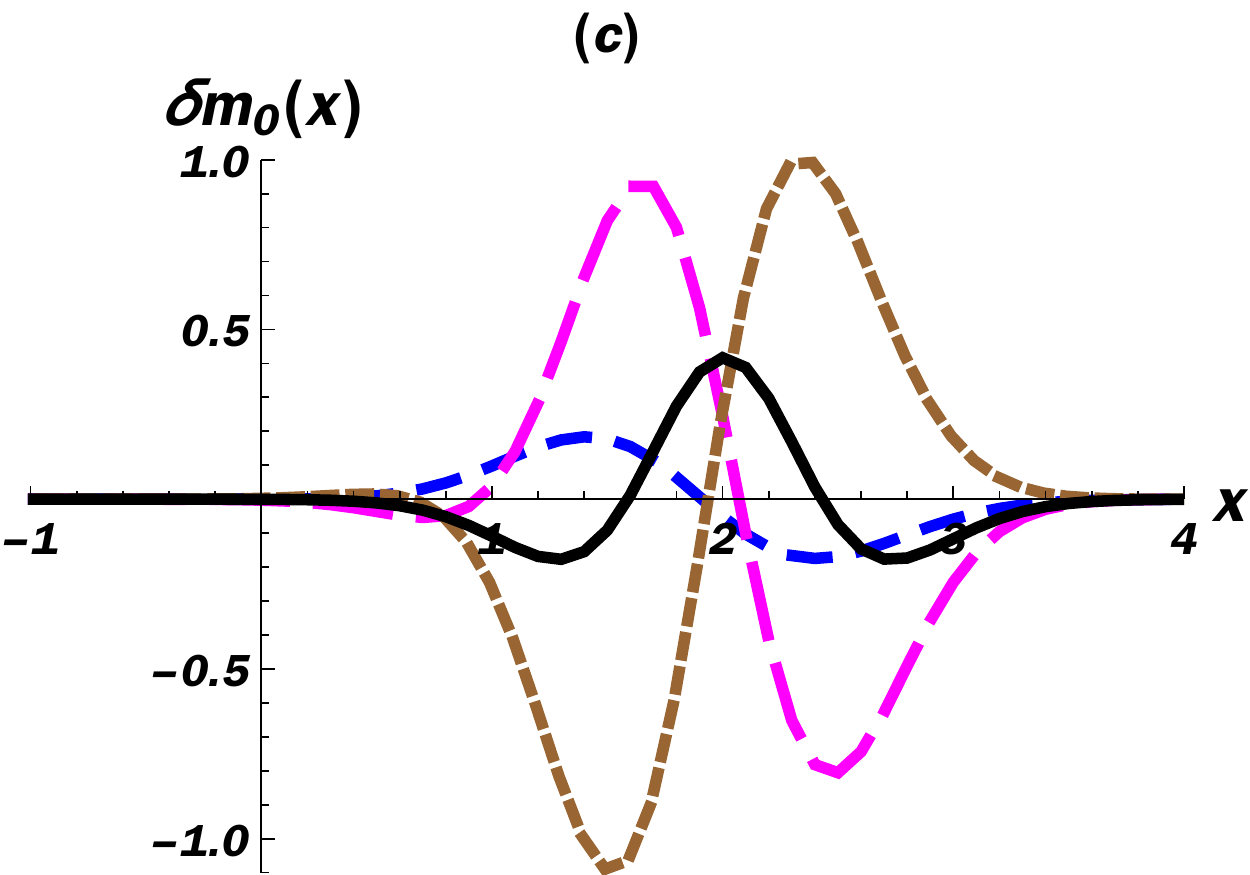}
\caption{Short time evolution of a density of agent in an harmonic
  potential in the limit of weak interactions (the units of time and
  length are the one set by the harmonics oscillator, i.e.\ $\omega
  =1$ and $l_0 =1$).  (a) Initial density of agents and its short
  time evolution. Solid black: Initial density of agent $m_0(x)$,
  chosen here as a Gaussian of width $\Sigma= .5$ centered at $x_0
  =2$. Solid gray: Short time evolution $m_0(x) + \Delta t \partial_t
  m^{(0)}_0(x)$ in the non interacting approximation ($\Delta t=0.1$).
  Dashed black: Short time evolution $m_0(x) + \Delta t (\partial_t
  m^{(0)}_0(x) + (g/\mu\sigma^2) \partial_t m^{(1)}_0(x))$ including
  the first order corrections associated with interactions ( the
  ``small parameter'' $(g/\mu\sigma^2)$ has been set to one to enhance
  visibility). (b) Time derivative of the density.  Solid gray:
  non interacting term $\partial_t m^{(0)}_0(x)$.  Solid black first
  order interaction term $\partial_t m^{(1)}_0(x)$. (c)
  The three contributions to the interacting term
  (Eq. \eqref{eq:intdec}).  Dashed (blue online) contribution
   $\partial_t m^{(e)}_0(x))$ associated
  with the perturbation of the ergodic state. Long-dashed (magenta
  online) contribution  $\partial_t
    m^{(a)}_0(x))$ associated with anticipations. Dotted (brown
  online) contribution  $\partial_t
    m^{(r)}_0(x))$ associated with the 
    classical (retarded) effects of interactions. Solid line: total
  contribution (same curve as in (b)).}
\label{fig:shorttime}
\end{figure}

A few remarks are in order.  In Fig.~\ref{fig:shorttime}a, the short
time free evolution of the density is what is expected, i.e.\ a motion
toward the maxima $x=0$ of the ``utility'' function $U_0(x)$; the
corrections due to the interactions are indeed quite small in the case
presented here.  This smallness is due in part to the fact that the
three contributions in \eqref{eq:intdec} ``push'' in different
directions and compensate one another for a large part (see
Fig.~\ref{fig:shorttime}c).  The terms $\partial_t m^{(e)}_0(x)$
associated with the modification of the {\em ergodic state} and
$\partial_t m^{(a)}_0(x)$ associated with {\em anticipations} tend to
accelerate the motion, when the retarded contribution $\partial_t
m^{(r)}_0(x)$ tends to slow it down (as the interactions make the
initial mean location slightly more favorable than in the free case).
Beyond a tendency to make the distribution slightly more narrow, the
net effect of interactions {\em for this particular example} is thus
to effectively slightly slow the motion of the group.  We stress
however that for the example considered here the width of the
initial density has been chosen  slightly smaller than the one of
the ergodic state, and their mean positions not too far away.
Other choices may have led to a stronger influence
of the anticipations.

\section{Conclusion}
\label{sec:conclusion}

In this paper, after a general  introduction to Mean Field Games
  in the form in which they have been introduced orginally by Lasry
  and Lions, \cite{LasryLions2006-1,LasryLions2006-2,LasryLions2007},
  and a bird's eye survey of the recent mathematical development in
  that field, we have considered in details a class of mean field game models,
referred to here as ``quadratic'' MFG. Such models describe the
collective behavior of a large number of agents, whose individual
dynamics follow a controlled linear Langevin equation,  when the control
derives from the minimization of a quadratic cost functional.

As we have emphasized, there exist a formal,
but deep, relationship between the MFG equations describing these
models and the nonlinear Schr\"odinger equation. Our main purpose was
to explore this relationship and its implications on the structure of
the solutions of MFG problems.

Indeed, the nonlinear Schr\"odinger equation has a very long history
in physics, and many tools and approximation schemes have been
developed along the years to analyze its properties in different
parameter regimes.  Using the connection between MFG and NLS, we have
shown that it is possible to adapt some standard tools from Quantum
Mechanics (Ehrenfest relations, perturbative expansions), or to rely
on concepts and techniques more specific to the nonlinear
Schr\"odinger equation (here the concept of soliton, the existence of
an action from which the equations of motion derive, and the related
variational approaches) and that it may indeed lead to a sharp control
over the behavior of quadratic MFG models in various regimes. In
particular we have introduced variational methods that are well
adapted in the limit of strong interaction, while on the opposite,
weak interaction limit, a perturbation theory can be developed.  A few
(partly new) exact results have been derived along the way.

In this paper, we have mostly limited ourselves to an introduction of
these methods in the context of Mean Field Games, and illustrated them
on a few simple examples, but it is already clear that this is very
far from exhausting the possible applications of the connection with
the nonlinear Schr\"odinger equation.

To start with, preliminary results show that other tools developed in
the context of the nonlinear Schr\"odinger equations, from simple ones
such as the Thomas Fermi approximation to more sophisticated one
related to inverse scattering methods, can be used to analyze further
the behavior of ``paradigmatic'' population models in the same spirit
as what have been done here.  More elaborated models, including for
instance the presence of two or more groups of agents with different
behaviors, or other specific modifications can also certainly be
analyzed following the same approach.

Obviously, the class of models considered here forms only a small
  subset of all possible Mean Field Games, but is already large enough
  to contain non trivial examples for which no explicit exact
  solutions are to be found. It includes both potential models (for
  which an action functional can be defined) and non-potential ones
  (for which this is not possible), as well as monotone (repulsive
  interactions in the model treated here ) and non-monotone ones (here
  the attractive case)\footnote{From a mathematical point of view, MFG
    which are both monotone and potential are somewhat easier to
    control.  Here we have focused on examples which can be associated
    to an action functional, but kept with the richer, non-monotone
    case.}.  Therefore, it seems to us relatively clear that, within a
  relatively short amount of time, the connection between quadratic
  Mean Field Games and the nonlinear Schr\"odinger equation will
  provide a good understanding of a large class of Mean Field Game
  problems.  We believe that quadratic MFG are at some level
  representative of a much more general class of Mean Field Games and
  expect that the connection outlined here will contribute
  significantly to a better understanding of general MFG models, as
  well as extending the field of their possible applications. In
    this process, we are confident that the physics community, with
    its specific knowledge and  point of view, can, and
    should, play an significant role.  We hope that this paper will
    be instrumental in this respect.

\bigskip

{\bf Acknowledgments:} This research has been conducted as part of the project
Labex MME-DII (Funded by the Agence Nationale pour la
Recherche, Grant No. ANR11-LBX-0023-01).

\appendix

\renewcommand{\theequation}{A.\arabic{equation}}

\section{Derivation of the Hamilton-Jacobi-Bellman equation}
\label{app:HJB}

The Hamilton-Jacobi-Bellman equation is one of the basic tool of
optimal control and is discussed in details in many
textbooks, such as for instance \cite{BertsekasBook}.  As it might be less
familiar to physicists, we provide a brief sketch of its derivation
below.

Starting from the definition Eq.~\eqref{eq:value} of the value
function $u(\bx,t)$, the principle of dynamic programming consists
in splitting the optimization in two parts, the first one for the
infinitesimal time  interval $[t,t+dt]$, and the second for all times
beyond this.  This reads
\begin{equation} \label{eq:LP}
 u(\bx,t) = \min_{{\bf a(t)}} 
\llangle \displaystyle \left[ L({\bf x}(t),{\bf a}(t)) -
  \tilde V[m_{t}]({\bf x}(t)) \right] \, + \, 
u(\bx + d\bX, t + dt)  \rrangle_{\rm   noise} \; , 
\end{equation}
with $d\bX$ given by the Langevin equation~\eqref{eq:Langevin}.
Using It\^o lemma we then have
\[ 
\llangle [u(\bx + d\bX, t + dt)]\rrangle_{\rm   noise} = u(\bx, t) + 
\left[\partial_t u(\bx,t)  + {\bf a} \cdot \partial_\bx u(\bx,t) +
\frac{\sigma^2}{2} \partial^2_{\bx,\bx} u(\bx,t) \right] dt \; ,
\]
which, inserted in Eq.~\eqref{eq:LP} yields the Hamilton-Jacobi-Bellman equation
\begin{equation}
\partial_t u_t(\bx) +H(\bx, \partial_\bx u_t(\bx))+
         \frac{\sigma^2}{2}  \partial_{\bx\bx} u_t(\bx) = 
        \tilde V[m_t](\bx) \; ,
 \end{equation}
 where
 $H(\bx,\bp) \equiv \inf_{\pmb \alpha} \left( L(\bx,\pmb \alpha)+\bp
   .\pmb \alpha) \right)$.

 The boundary condition $u(\bx,t\!=\!T) = c_T(\bx)$ is then
   obtained by noting that at the end of the optimization interval,
   there is no control variable on which the agent can act to optimize
   its cost, and therefore the utility function is just given by the
   final cost $c_T$.

This completely solve the optimization problem for the cost
  function when one assumes that the value function is twice
  differential. In more general settings, weak or viscosity solutions
  for the HJB equation has to be considered instead
  \cite{LionsMenaldi1982}.

\renewcommand{\theequation}{B.\arabic{equation}}

\section{ Solutions of the generalized Gross-Pitaevskii equation in one dimension}
\label{app:GP}

In one dimension and in the absence of external potential, $U_0 =
  0$, the ergodic problem considered in subsection \ref{sec:zero-U}
  reduces to the following generalized Gross-Pitaevskii equation
  \eqref{eq:GP}: 
\begin{equation}
	\frac{\mu \sigma^4}{2}\partial^2_{xx} \psi^e(x) + g (\psi^e(x))^{(2 \alpha+1)} = - \lambda \psi^e(x) \;.
\end{equation}
Integrating once gives
\begin{equation}\label{eq:GP2}
	\frac{\mu \sigma^4}{4} (\partial_x \psi^e(x))^2+\frac{g}{2
          \alpha+2}(\psi^e(x))^{(2
          \alpha+2)}+\frac{\lambda}{2}(\psi^e(x))^2= 0 \; ,
\end{equation}
where the integration constant (the right hand side of Equation
  \eqref{eq:GP2}) is set to zero  since a solution
associated with the minimum value for $\lambda^e$ has to decay
 to zero at infinity:
\begin{equation*}
	\lim_{|x| \to \infty} \psi^e(x) = \lim_{|x| \to \infty} \partial_x \psi^e(x)=0 \;.
\end{equation*}
The function $\psi^e(x)$ need to have (at least) a nonzero maximum $\psi_M$, 
 which value is thus the unique positive solution of 
\begin{equation}
	\frac{g}{2
          \alpha+2}\psi_M^{(2\alpha+2)} + \frac{\lambda}{2}\psi_M^2=0
        \; ,
\end{equation}
which imposes $\lambda^e <0$ and ,
\begin{equation}\label{eq:psim-B}
	\psi_M=\left(\frac{- (\alpha+1) \lambda^e}{g}
        \right)^{1/2\alpha} \; . 
\end{equation}
Defining  a characteristic length $\eta_\alpha$ as
\begin{equation}\label{eq:eta-B}
\eta_\alpha= \sqrt{\frac{\mu\sigma^4}{-2 \alpha^2\lambda^e}} \; , 
\end{equation}
equation \eqref{eq:GP2} can be reduced in the form
\begin{equation}
\alpha^2\,\eta_\alpha^2  \bigl(\frac{\partial_x
  \psi^e(x)}{\psi_M}\bigr)^2 - \bigl(\frac{ \psi^e(x)}{\psi_M}\bigr)^2
\Bigl( 1 - \bigl(\frac{ \psi^e(x)}{\psi_M}\bigr)^{2 \alpha} \Bigr) = 0
\; ,
\end{equation}
which can be readily integrated as 
\begin{equation}
\psi^e(x)= \psi_M\;\bigl[\cosh( \frac{x-x_0}{\eta_\alpha})\bigr]^{-
  \frac{1}{\alpha}} \; .
\end{equation}

Finally, the value of $\lambda$ is   fixed through the normalization condition:
	\begin{equation*}
		\int dx \,(\psi^e(x))^2  =1 \; .
	\end{equation*}
Setting
\begin{equation}
I_\alpha=
\int_{0}^{+\infty}(\cosh(x))^{-\frac{2}{\alpha}} dx =
4^{\frac{1-\alpha}{\alpha}}\frac{\Gamma(\frac{1}{\alpha})^2}{\Gamma(\frac{2}{\alpha})},
\end{equation}
we get
\begin{equation}
\lambda^e = -\bigl(\frac{g}{\alpha+1}\bigr)^{\frac{2}{2-\alpha}}
	\bigl(\frac{\alpha}{\sqrt{2 \mu\sigma^4} I_\alpha}\bigr)^{\frac{2\alpha}{2-\alpha}} 
= -\frac{1}{4} \; \Bigl(
\frac{\Gamma(\frac{2}{\alpha})}{\Gamma(\frac{1}{\alpha})^2}\Bigr)^\frac{2\alpha}{2-\alpha}
\;
\bigl(\frac{g}{\alpha+1}\bigr)^\frac{2}{2-\alpha}\;
\Bigl(\frac{2\alpha^2}{\mu\sigma^4}\Bigr)^\frac{\alpha}{2-\alpha} \; ,
\end{equation}	
and the expression of the characteristic length \eqref{eq:eta-B} becomes
\begin{equation}
\eta_\alpha=2\;
\Bigl(\frac{\Gamma(\frac{1}{\alpha})^2}{\Gamma(\frac{2}{\alpha})}\Bigr)^{\frac{\alpha}{2-\alpha}}\,
\Bigl(\frac{\alpha+1}{2\alpha^2}\,\frac{\mu\sigma^4}{g}\Bigr)^\frac{1}{2-\alpha}
\; .
\end{equation}


\renewcommand{\theequation}{C.\arabic{equation}}

\section{Gaussian variational ansatz}
\label{app:variational}
In this appendix, we provide some of the intermediate results that has
been used when discussing the variational approach used in
sections~\ref{sec:strong} and \ref{sec:more-complicated}.

We consider a MFG model with $d$-dimensional state space, non-linear local interactions and an external potential as 
in \eqref{eq:NLinPot}.

The action Eq.~(\ref{eq:S}) can be written as 
\[
S[\Phi,\G] = \int_0^T dt L(t) \; ,
\]
 with a Lagrangian $L(t) = L_\tau +( E_{\rm kin} + E_{\rm pot} +
E_{\rm int})$ where
\begin{align*}
L_\tau(t) &= - \frac{\mu \sigma^2}{2} \int  d\bx 
\Phi(\bx,t) \bigl((\partial_t \Gamma(\bx,t)) -  (\partial_t \Phi(\bx,t))  \Gamma(\bx,t)  \Bigr)  \; , \\
E_{\rm kin}(t) & = \frac{1}{2\mu} \langle \hat \Pi^2\rangle_t = -
\frac{\mu \sigma^4}{2} \int  d\bx 
\nabla \Phi(\bx,t)  \cdot  \nabla \Gamma(\bx,t) \; ,
 \\
E_{\rm pot} (t) & =  \langle U_0\rangle_t=   \int  d\bx \, \Phi(\bx,t)
\,U_0(\bx)\, \Gamma(\bx,t)\; ,   \\
E_{\rm int} (t) &=   \int  d\bx \, F[ \Phi(\bx,t)\Gamma(\bx,t) ]  
=  \frac{g}{\alpha+1} \int  [  \Phi(\bx,t)\Gamma(\bx,t) 
]^{\alpha+1}d\bx \; .
\end{align*}

We hereafter develop a variational ansatz by minimizing this action on a restricted class of functions $(\Phi(\bx,t),\Gamma(\bx,t))$
defined as in \eqref{eq:AnsatzGamma}-\eqref{eq:AnsatzPhi} that we recall here for convenience:
\begin{align}
\Phi(\bx,t) & =  \exp \Bigl\{ \frac{-\gamma_t + \bP_t \cdot \bx
   }{\mu\sigma^2}  \Bigr\}
\prod_{\nu=1}^d \left[
\frac{1}{\bigl(2\pi (\Sigma^\nu_t)^2\bigr)^{1/4}}
\exp  \Bigl\{-\frac{(x^\nu- X^\nu_t)^2}{ (2\Sigma^\nu_t)^2} 
(1 - \frac{\Lambda^\nu_t}{\mu \sigma^2} )\Bigr\}
\right] \; .
 \label{eq:AnsatzPhi-A} \\
 \G(\bx,t) &= \exp \Bigl\{\frac{+\gamma_t - \bP_t \cdot \bx}{\mu
  \sigma^2} \Bigr\} 
\prod_{\nu=1}^d \left[
\frac{1}{\bigl(2\pi (\Sigma^\nu_t)^2\bigr)^{1/4}}
\exp \Bigl\{-\frac{(x^\nu- X^\nu_t)^2}{ (2\Sigma^\nu_t)^2} 
(1+\frac{\Lambda^\nu_t}{\mu \sigma^2} )\Bigr\}
\right]  \; ,
   \label{eq:AnsatzGamma-A}
\end{align}

\subsection*{Computation of the Lagrangian}
We obtain for the  first two  terms 
\begin{align}
L_\tau(t) &=  - \dot \gamma_t 
+\sum_{\nu =1}^d \frac{\dot \Lambda_t^\nu}{4}
+  \dot \bP_t \cdot\bX_t \,
  - \sum_{\nu =1}^d \frac{ \Lambda_t^\nu  \dot \Sigma_t^\nu}{2 \Sigma_t^\nu}
         \label{eq:Ltau}\\
E_{\rm kin}(t) &=   \frac{\bP_t^2}{2\mu} \,  + 
\sum_{\nu =1}^d
  \frac{(\Lambda_t^\nu)^2 - \mu^2 \sigma^4}{8\mu (\Sigma_t^\nu)^2}  \; ,
\end{align}
and we choose from now on
\begin{equation}
\gamma_t=\gamma_0+\sum_{\nu =1}^d \frac{\Lambda_t^\nu}{4} \; ,
\end{equation}
so that the first two terms in \eqref{eq:Ltau} cancel.

The computation of the potential energy would requires 
the external potential
$U_0$ to be  given. However, since  $E_{\rm pot}$  depends only on
$m(\bx,t) = \G(\bx,t) \Phi(\bx,t)$, and not separately on both $\G(\bx,t)$ and
$\Phi(\bx,t)$  it depends on $\bX_t$ and
$\mathbf{\Sigma}_t$ only.  Furthermore, using a Taylor expansion of $U_0(\bx)$
around $\bX_t$, we get 
\begin{equation} \label{eq:Lpot} 
E_{\rm pot}(t) =   U_0(\bX_t) + \frac{1}{2} \sum_{\nu=1}^d (\Sigma_t^\nu)^2 
 \partial^2_{\nu\nu} U_0(\bX)  + O(\|\mathbf{\Sigma}_t\|^4) \; .
\end{equation}

Finally, the interaction energy in the variational ansatz reads 
\begin{align}
E_{\rm int} (t) &=  \frac{g}{\alpha+1} \int  [ \Phi(\bx,t)  \Gamma(\bx,t)]^{\alpha+1}d\bx\nonumber \\
&=  \frac{g}{\alpha+1} \prod_{\alpha=1}^d \left[ \frac{1}{\sqrt{2\pi (\Sigma_t^\nu)^2}}
\int dx \exp\Bigl\{\frac{(\alpha+1) (x-\bX^\nu_t)^2}{2 (\Sigma^\nu_t)^2}\Bigr\}\right]\nonumber\\
&=\frac{g}{\alpha+1} \frac{1}{((\alpha+1)(2\pi)^\alpha)^{d/2}}
\prod_{\nu=1}^d \Bigl(\frac{1}{\Sigma^\nu_t}\Bigr)^\alpha \; .
\end{align}

The two pairs of variables, $(\bX,\bP)$ and
$(\mathbf{\Sigma},\mathbf{\Lambda})$ are coupled only through the
potential energy \eqref{eq:Lpot}, as a consequence of the curvature
of $U_0$ on the scale $\|\mathbf{\Sigma}_t\|$. Assuming that
these corrections are negligible, the two pairs
  decouple and evolve independently. The motion of the center of mass
  follows a reduced dynamics in the external potential $U_0$:
\begin{align*}
\dot\bX_t&= \frac{P_t}{\mu} \; ,\\
\dot\bP_t&= -\mathbf{\nabla} U_0(\bX)\; ,
\end{align*}
and the total energy of the center of mass $\frac{\bP^2}{2\mu}+
U_0(\bX)$ is separately conserved.  On the other hand, the
  dynamics in the center of mass,
$(\mathbf{\Sigma}_t,\mathbf{\Lambda}_t)$ is
 governed by the reduced action $\tilde
S(\mathbf{\Sigma},\mathbf{\Lambda}) = \int_0^T dt \tilde L(t)$ with
\begin{align}
\tilde L(t) &= \tilde L_\tau(t) + \tilde E_{\rm
  tot}(t) \; , \label{eq:Ltilde} \\
\tilde L_\tau(t) &= 
- \sum_{\nu =1}^d \frac{\Lambda^\nu_t \dot\Sigma^\nu_t}{2\Sigma^\nu_t}
\; , \\
\tilde E_{\rm tot}(t) &=  \sum_{\nu =1}^d \frac{(\Lambda_t^\nu)^2-\mu^2\sigma^4}{8\mu(\Sigma^\nu_t)^2}
+ \frac{ g }{\alpha+1} \prod_{\nu =1}^d 
\left[\frac{1}{\sqrt{\alpha+1} (2\pi)^{\alpha/2}}
  \Bigl(\frac{1}{\Sigma^\nu_t}\Bigr)^\alpha\right] \; . \label{eq:rEtot}
\end{align}

The $2d$ equations of motion
\eqref{eq:Sigma-dot}--\eqref{eq:Lambda-dot} are obtained by computing
the variations of the action along the trajectories and equating them
to zero. They read
\begin{align}\label{eq:Sigma_d_alpha}
\dot \Sigma^\nu_t &= \frac{\Lambda^\nu_t}{2\mu\Sigma^\nu_t}\\
\dot \Lambda^\nu_t &=\frac{(\Lambda_t^\nu)^2-\mu^2\sigma^4}{2\mu(\Sigma^\nu_t)^2}
+\frac{2 g\alpha }{\alpha+1} \prod_{\rho =1}^d 
\left[\frac{1}{\sqrt{\alpha+1} (2\pi)^{\alpha/2}}
  \Bigl(\frac{1}{\Sigma^\rho_t}\Bigr)^\alpha\right] \; .
\label{eq:Lambda_d_alpha}
\end{align}

These equations admit one stationary point  at  which the
  variances and position-momentum correlators are the
same in all directions, namely
\begin{align}
\Lambda^\nu_*&=\Lambda_*=0\\
\Sigma^\nu_*&=\Sigma_*=\left[\frac{4\alpha}{\alpha+1}\Bigl(\frac{1}{(\alpha+1)(2\pi)^\alpha}\Bigr)^{d/2}
  \frac{g}{\mu\sigma^4}\right]^{-1/(2-\alpha
  d)} \; , \label{eq:Sigma*_d_alphaB} 
\end{align}
while the total energy at the stationary point $(\mathbf{\Sigma}_*,\mathbf{\Lambda}_*) $ is
\begin{equation}\label{eq:Etot*_d_alpha}
\tilde E_{\rm tot}^*=\frac{2-\alpha d}{8\alpha}
\frac{\mu\sigma^4}{\Sigma_*} \; .
\end{equation}
Note also that the reduced kinetic and interaction energy are respectively 
\begin{align}
\tilde E_{\rm kin}^*=-\frac{\alpha d}{2-\alpha d} \tilde E_{\rm tot}^*
\; , \\
\tilde E_{\rm int}^*=\frac{2}{2-\alpha d} \tilde E_{\rm tot}^* \; .
\end{align}
Eliminating $\mathbf{\Lambda}_t$ and its derivatives from the
evolution equations \eqref{eq:Sigma_d_alpha}-\eqref{eq:Lambda_d_alpha}
lead to a set of second order  
coupled equations in the variables $\Sigma^\nu_t$ only:
\begin{equation} 
\ddot \Sigma^\nu_t= -\frac{\sigma^4}{8 \Sigma^\nu_t}
+\frac{\alpha}{\alpha+1} \prod_{\rho =1}^d 
\left[\frac{1}{\sqrt{\alpha+1} (2\pi)^{\alpha/2}}
  \Bigl(\frac{1}{\Sigma^\rho_t}\Bigr)^\alpha\right] \frac{g}{\mu
  \Sigma^\nu_t} \; .
\end{equation}
Using expressions Eqs.~\eqref{eq:Sigma*_d_alphaB}-\eqref{eq:Etot*_d_alpha}, and introducing the reduced variables
$q_t^\nu= \Sigma^\nu_t/\Sigma_*$, we get the simpler expression:
\begin{equation} 
\ddot q^\nu_t = \frac{\alpha }{2-\alpha d}\frac{2\tilde E_{\rm tot}^*}{ \mu\Sigma_*^2}\left[-\Bigl(\frac{1}{ q^\nu_t}\Bigr) ^3 
+\frac{1}{ q^\nu_t} \prod_{\rho =1}^d
\Bigl(\frac{1}{q^\rho_t}\Bigr)^\alpha\right] \; ,
\end{equation} 
and in these variables, conservation of energy reads
\begin{equation}\label{eq:Csrv_d_alphaA}
\sum_{\nu=1}^d (\dot q^\nu_t)^2=\frac{2\tilde E_{\rm tot}^*}{ \mu\Sigma_*^2}\left[ \frac{\alpha }{2-\alpha d}\sum_{\nu=1}^d \Bigl(\frac{1}{ q^\nu_t}\Bigr) ^2 
-\frac{2 }{2-\alpha d} \prod_{\nu =1}^d
\Bigl(\frac{1}{q^\nu_t}\Bigr)^\alpha\right] 
+\frac{2\tilde E_{\rm tot}}{ \mu\Sigma_*^2} \; .
\end{equation}

\subsection*{Invariant manifolds in one dimension}

The fixed point $(\Sigma_*,\Lambda_*)$ of the dynamical system
Eqs.~\eqref{eq:Sigma-dot}-\eqref{eq:Lambda-dot} is unstable for
$\alpha \in ]0,2[$ and the associated soliton is stable.  Along the
stable and unstable manifolds associated with the stationary point, the
total energy $\tilde E_{\rm tot}=\tilde E^*_{\rm tot}$ and Equation
\eqref{eq:Csrv_d_alphaA} reads
\begin{equation}\label{eq:Csrv_d_alphaB}
\sum_{\nu=1}^d (\dot q^\nu_t)^2=\frac{2\tilde E_{\rm tot}^*}{
  \mu\Sigma_*^2}\left[ \frac{\alpha }{2-\alpha d}\sum_{\nu=1}^d
  \Bigl(\frac{1}{ q^\nu_t}\Bigr)^2  
-\frac{2 }{2-\alpha d} \prod_{\nu =1}^d
\Bigl(\frac{1}{q^\nu_t}\Bigr)^\alpha+1\right] \; .
\end{equation} 
In one space dimension, the expressions for the width
\eqref{eq:Sigma*_d_alphaB} and the total energy
\eqref{eq:Etot*_d_alpha} at the stationary point simplify: 
\begin{align}
\Sigma_*|_{d=1}&=\left[\frac{(2\pi)^{\alpha/2}(\alpha+1)^{3/2}}{4\alpha}
  \frac{\mu\sigma^4}{g}\right]^{1/(2-\alpha)} \; , \\
\tilde E_{\rm tot}^*|_{d=1}&=\frac{2-\alpha }{8\alpha}
\frac{\mu\sigma^4}{\Sigma_*} \; , 
\end{align}
and the equation \eqref{eq:Csrv_d_alphaB} can be integrated, giving
the equation for the stable and unstable manifolds.  Introducing the
function $F(q)$ as the integral
\begin{equation}
F(q)=
\begin{cases}
\displaystyle \int_0^{q}
\frac{\mathrm{d}v}{(v-\frac{v^{1-\alpha/2}}{1-\alpha/2} -
  (1-\frac{1}{1-\alpha/2}))^{1/2}} &\text{ if } q<1 \; ,\\
\displaystyle - \int_q^\infty
\frac{\mathrm{d}v}{(v-\frac{v^{1-\alpha/2}}{1-\alpha/2} -
  (1-\frac{1}{1-\alpha/2}))^{1/2}} & \text{ if } q>1 \; .
\end{cases}
\end{equation}
The equation for the stable manifold reads
\begin{equation}
	F(q_t)-F(q_0)=\frac{t}{\tau_*}  \; ,
\end{equation}
while the  equation for the unstable manifold is
\begin{equation}
	F(q_t)-F(q_T)=\frac{T-t}{\tau_*}  \; ,
\end{equation}
where the characteristic time $\tau_*$ is 
\begin{equation}
	\tau_*= \sqrt{\frac{\mu\Sigma_*^2}{2 \tilde E_{\rm tot}^*}}
\end{equation}
Note that the function $F(q)$ behaves as
$F(q)\approx\frac{1}{\sqrt{\alpha}}\log{|1-q|}$ when $q\approx 1$, as
$F(q)\approx\frac{2-\alpha}{2\alpha} q^2$ for $q\ll 1$ and like
$F(q)\approx q$ for $q\gg 1$ .


\renewcommand{\theequation}{D.\arabic{equation}}

\section{Propagator for the harmonic oscillator}
\label{app:perturbative}

In this appendix, we give a brief derivation of the expression  for the 
propagator  Eq.~\eqref{eq:prop-explicit} corresponding to  an harmonic
potential. This formula  is a rather classical result and various
derivations can be found in \cite{WATSON}; the following one is given
here for completeness.  We  first set the length unit so that $\ell_0 \equiv 1$.

We consider the  n$^{\rm th}$ eigenfunction of the harmonic oscillator $\psi_n$ \eqref{eq:varphi-n} 
and introduce its Wigner transform as
\[
W_n(x,p) \equiv \int dy \, e^{-i y p} \, \psi_n(x-y/2) \psi_n(x+y/2) \; ,
\]
 We get 
\[ 
W_n(x,p) = 2 (-1)^n \exp \left[ -(x^2+p^2) \right] L_n \left(2
  (x^2+p^2) \right) \; , \]
 where $L_n \equiv \frac{e^x}{n !} \frac{d^n}{dx^n} (x^n e^{-x})$ is the $n^{\rm th}$ 
Laguerre polynomial.

The Wigner transform of the propagator \eqref{eq:prop-explicit} therefore reads 
\begin{align*}
W_{\hat G}(x,p,t) &\equiv \sum_n \exp\left(- \frac{t \lambda_n}{\mu
    \sigma^2} \right) W_n(x,p)  \\ 
& = 2 \exp \left[ -(x^2+p^2)  \right] \exp \left[ - \frac{\omega t}{2}
\right] \sum_n (-1)^n \exp \left[ - \frac{n \omega t}{2}
\right] L_n \left(2   (x^2+p^2) \right) \\
& =
\frac{\exp \left[- (x^2+p^2)  \tanh \left( \frac{\omega t}{2} \right)
  \right ]}{\cosh \left(\frac{\omega t}{2} \right)}
\; ,
\end{align*}
where, in order to get the last line, we have used that 
\[ \sum_n L_n(z) \xi^n = \exp [-z \xi/(1-\xi) ] /
(1-\xi)\]
with $z= 2 (x^2+p^2)$ and $\xi = - e^{-\omega t}$.

Through an inverse Fourier transform, the propagator
\eqref{eq:prop-explicit} is now expressed as
\begin{align*}
G(x,x',t) & \equiv \int \frac{dp}{2\pi} e^{i p(x'-x)} W_{\hat
  G}({\scriptstyle \frac{1}{2}}(x+x'),p,t) \\ 
& = \frac{1}{\sqrt{2\pi \sinh(\omega t)}} \exp \left[
-\frac{(x^2+x'^2)\cosh(\omega t) -2xx'}{\sinh(\omega t)} \right]
\; .
\end{align*}
Reinserting by homogeneity the dependence on the length scale $\ell_0$
leads to Eq.~\eqref{eq:prop-explicit}.

\bibliographystyle{plain}
\bibliography{Biblio-2018-07-03}

\end{document}